\documentclass[aps,prl,reprint,nofootinbib,superscriptaddress]{revtex4-2} 
\usepackage{commons}
\usepackage{CJK}

\usepackage{graphicx}

\usepackage{dcolumn}

\usepackage{bm}
\usepackage[normalem]{ulem}
\usepackage{color}
\usepackage{amsmath}
\usepackage{amssymb}
\usepackage{lipsum}
\usepackage{bm}
\usepackage{xcolor}
\usepackage{graphicx}
\usepackage{verbatim}
\usepackage[T1]{fontenc}
\usepackage{hyperref}
\hypersetup{colorlinks=true,linkcolor=blue}
\usepackage{orcidlink}

\usepackage{pgffor}
\usepackage{pdfpages}
\makeatletter
\AtBeginDocument{\let\LS@rot\@undefined}
\makeatother

\newcommand{\SR}[1]{\textcolor{black}{#1}}

\newcommand{\MOD}[1]{\textcolor{black}{#1}}

\begin{document}
\title{{Superdiffusion and antidiffusion in an aligned active suspension}}
\author{Lokrshi Prawar Dadhichi$^*$}
\email{lpdadhichi@gmail.com}
\affiliation{Institute for Theoretical Physics, Leipzig University, 04103 Leipzig, Germany}
\author{Suvendra K. Sahoo$^*$
}
\email{suvendrak@iisc.ac.in}
\affiliation{Centre for Condensed Matter Theory, Department of Physics, Indian Institute of Science, Bangalore 560 012, India}
\author{K. Vijay Kumar
}
\email{vijaykumar@icts.res.in}
\affiliation{International Centre for Theoretical Sciences,
Tata Institute of Fundamental Research, Bengaluru 560 089, India}
\author{Sriram Ramaswamy}
\email{sriram@iisc.ac.in}
\affiliation{Centre for Condensed Matter Theory, Department of Physics, Indian Institute of Science, Bangalore 560 012, India}
\affiliation{International Centre for Theoretical Sciences,
Tata Institute of Fundamental Research, Bengaluru 560 089, India}

\begin{abstract}
%
We show theoretically that an imposed uniaxial anisotropy leads to new universality classes for the dynamics of active particles suspended in a viscous fluid. In the homogeneous state, their concentration relaxes superdiffusively, stirred by the long-ranged flows generated by its own fluctuations, as confirmed by our numerical simulations. Increasing activity leads to an anisotropic diffusive instability, \MOD{and thus an original phase-separation mechanism,} driven \SR{by the interplay of active stresses with a 
particle current proportional to the local curvature of the suspension velocity profile}. 
\end{abstract}
\maketitle

\def\thefootnote{*}\footnotetext{These authors contributed equally to this work.}\def\thefootnote{\arabic{footnote}}

\begin{figure}
\includegraphics[width=0.98\columnwidth]{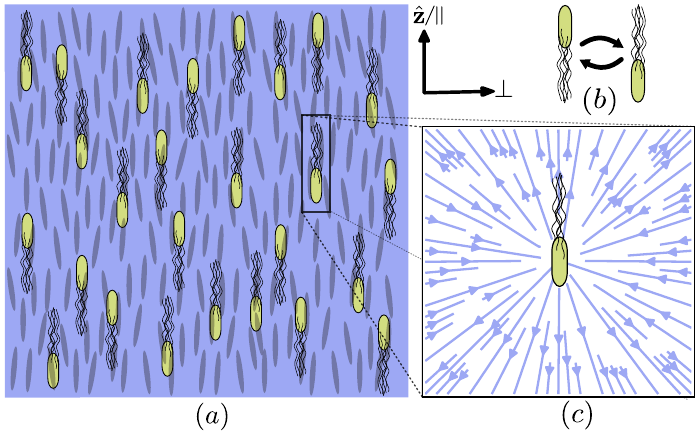}
\caption{(a) Schematic realization of (b) tumbling but aligned swimmers, in the form of bacteria homogeneously dispersed in a stiff nematic liquid crystal that aligns their force dipoles along its fixed $\hat{\bf z}$ axis.  (c) Far-field fluid velocity field due to a swimmer.}
\label{introfig}
\end{figure}

Active Matter, that is, living materials and their imitations, is made up of components that continually extract mechanical work from a fuel supply 
\cite{ramaswamy2010mechanics,bowick2022symmetry}. 
The dynamical equations governing active materials differ from their passive counterparts through physical fluxes arising from maintained chemical forces \cite{Prawar_Dadhichi_2018}. 
In the cases of greatest interest, these fluxes 
dominate at large length scales, resulting in dynamical and statistical properties qualitatively different from those of a time-reversal-invariant system with the same spatial symmetries. Much insight has emerged recently \cite{activeB+bubbly_PRX,activeB_NComm,tiribocchi2015active,cates2019active,cates2025active} from a focus on active currents and stresses built from scalar fields \cite{finlayson1969convective} rather than alignment or flocking. In this Letter we 
study such scalar active matter in a permanently anisotropic momentum-conserving fluid, described, like active \cite{singh2019hydrodynamically,tiribocchi2015active} or passive \cite{hohenberg1977theory} model H, by concentration and hydrodynamic velocity fields, but with a preferred axis. We discuss possible experimental realizations, e.g., Fig. \ref{introfig}, towards the end of the paper. %

\begin{figure}
    \includegraphics[width=0.65\columnwidth]{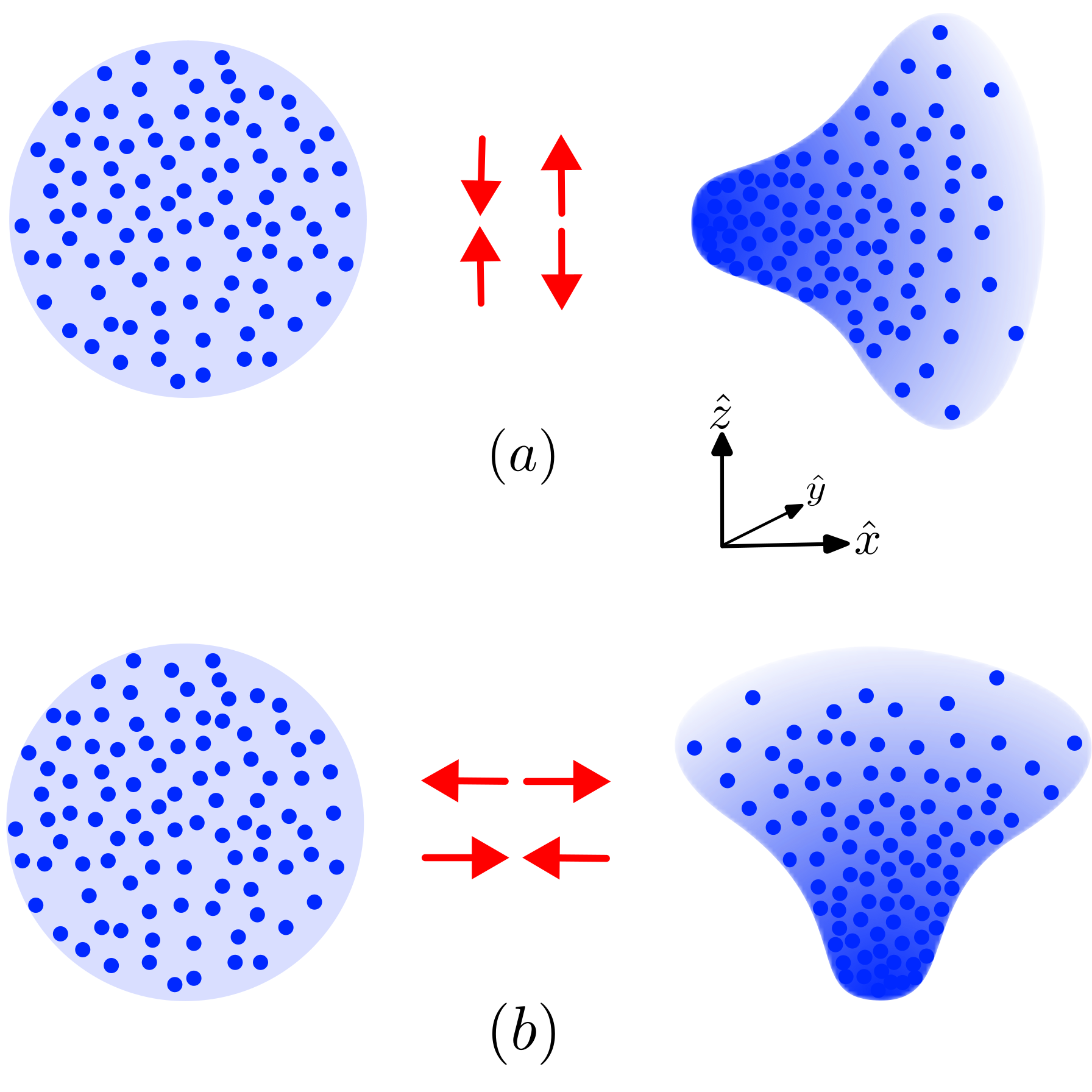}
    \caption{Left: 
    an initial isotropic configuration of concentration. Red arrows indicate the directions of ${\bf u}$ (a) when $\partial_\perp\partial_z u_z>0$, (b) when $\partial_z\partial_\perp u_\perp>0$, where $\perp=x$ or $y$; 
    right: 
    the concentration distribution after advection by the red-arrow flows.
    }
    \label{Flow}
\end{figure}

Here are our main results. 
The self-advection of concentration fluctuations by the long-ranged active flows they generate leads to a dynamic exponent $z = d/2$ for $d \le 4$: inhomogeneities on a length scale $L$ relax on a timescale $\propto L^{3/2}$ in $d=3$. Our numerical studies of Brownian force dipoles in a Stokesian fluid confirm this asymptotic superdiffusive scaling, and uncover an early-time ballistic dynamics that also emerges from our theory. 
In addition, we \SR{show that 
solute currents proportional to the curvature in the hydrodynamic velocity profile (\cref{Flow}) \cite{aubert1980macromolecules,callan2011hydrodynamics}, 
in concert with the active stresses, can drive a diffusive instability}, with a critical point at which the hydrodynamic interaction is strongly relevant. We thus uncover two hitherto unknown universality classes, describing the homogeneous phase and the onset of phase separation of uniaxial active model H. 

We now show how these results were obtained. We work with an active suspension, with solute concentration field $c({\bf r},t)$ and joint velocity field ${\bf u}({\bf r},t)$, as functions of position ${\bf r}$ and time $t$, and a macroscopic anisotropy defined by a distinguished $\hat{\bf z}$ direction, with $\hat{\bf z} \to -\hat{\bf z}$ symmetry and isotropy in the $\perp$ plane transverse to $\hat{\bf z}$ \textcolor{magenta}{\cite{dry_aniso_foot}}. We enforce overall incompressibility, i.e., total density $\rho=$ constant, as appropriate for the slow flows of interest here, so that $\nabla \cdot {\bf u} = 0$. The imposed, rather than spontaneous, alignment means that $c$ and ${\bf u}$ are the only slow variables, obeying the conservation laws $\partial_t c = - \nabla \cdot {\bf J}$ for number and $\partial_t (\rho {\bf u}) =  \nabla \cdot {\sigbold}$ for momentum, with current ${\bf J}$ and stress tensor $\sigbold$. To build our equations of motion, we need only ask what contributions to ${\bf J}$ and $\sigbold$, constructed from $c$ and ${\bf u}$ and their gradients, can arise in the \SR{presence of anisotropy 
that} compete with those already accounted for in (isotropic) active model H \cite{singh2019hydrodynamically,tiribocchi2015active}. A little reflection will show that there are precisely two contributions, at linear order in \SR{fields. One is a uniaxial active stress \cite{simha2002hydrodynamic} 
\be
    \label{newstress}
    \sigbold^a = -W \hat{\bf z} \hat{\bf z} c({\bf r},t)
\ee
which, given the fixed orientation, operates through inhomogeneities in the \textit{concentration}. The other, permitted in a passive system \MOD{but receiving important active contributions as well}, is a current} 
\begin{equation}
\begin{aligned}
\label{newcurrent}
\mathbf{J}_{\perp}^u =& \; a_1 \nabla_{\perp}\partial_{z}u_{z} + (a_2 \nabla_{\perp}^2 + a_3 \partial_z^2 + a_4  \nabla_{\perp}\nabla_{\perp}\cdot) \mathbf{u}_{\perp},\\
J_z^u =& \;b_1 \partial_{z}^{2}u_{z} + b_2 \partial_{z} \nabla_{\perp}\cdot\mathbf{u}_{\perp} + b_3 \nabla_{\perp}^2u_{z},
\end{aligned}
\end{equation}
\SR{which we will term Flow-Induced Migration (FIM). \MOD{The passive contribution to \eqref{newcurrent} 
can be viewed as an anisotropic variant of the migration due to inhomogeneous gradients discussed in \cite{aubert1980macromolecules,callan2011hydrodynamics} or indeed an extension of Fax\'{e}n's Law \cite{faxen1922WiderstandGegen,happel1983LowReynolds}}. 
Anisotropy is crucial: in the isotropic limit, ignoring the $c$-dependence of $a_i,b_i$ in \eqref{newcurrent}, $\nabla \cdot {\bf J}^u \propto \nabla^2 \nabla \cdot {\bf u}$ which does not contribute for an incompressible system.} Galilean invariance, incompressibility, and symmetry under inversion of, and rotation about, $\hat{\bf z}$ rule out contributions (apart from $c {\bf u}$) of ${\bf u}$ to the particle current at lower gradient order than \eqref{newcurrent}. 
Eqs. \eqref{newstress} and \eqref{newcurrent} lead to the equations of motion 
\begin{align}
    \label{ceq}
    (\partial_t + {\bf u}\cdot\nabla)c =  
    \MOD{- a \nabla_{\perp}^2 \partial_z u_z + b \partial_z^2 \nabla_{\perp} \cdot \mathbf{u}_{\perp}}  
    \nonumber \\
    + (D_\perp \nabla_\perp^2 + D_z \partial_z^2) c + \nabla\cdot \bf f + \ldots,
\end{align}
and 
\be
    \label{veq}
    \rho(\partial_t + {\bf u} \cdot \nabla) {\bf u} = \eta \nabla^2 {\bf u}  
    - {\nabla \mathcal{P}}
    - W \hat{\bf z} \partial_z c + \ldots,
\ee 
\MOD{so that the FIM current can be recast as ${\bf J}_{\perp}^u = a \nabla_{\perp} \partial_z u_z$, $J_z^u = - b \partial_z \nabla_{\perp} \cdot \mathbf{u}_{\perp}$, with $\{a_i\}, \{b_i\}$ in \eqref{newcurrent} packaged into two coefficients $a,b$.} \SR{The ellipsis denotes irrelevant contributions, including the Onsager partners \cite{callan2011hydrodynamics} of \eqref{newcurrent} in Eq. \eqref{veq}} \MOD{as well as terms present in active model H \cite{singh2019hydrodynamically,tiribocchi2015active} or their anisotropic variants.} $D_\perp$ and $D_z$ are bare diffusivities, $\eta$ is the shear viscosity, taken for simplicity to be isotropic, and the pressure $\mathcal{P}$ enforces incompressibility. 
In \cref{ceq}, we have allowed for number-conserving fluctuations in the form of a Gaussian random current $\mathbf{f}(\mathbf{r},t)$, white in space and time, with strengths $N_z$, $N_\perp$ respectively along and transverse to $\hat{\bf z}$. 
In \cref{veq} $W$ is positive (negative) for extensile or pusher (contractile or puller) swimmers \SR{aligned along $\hat{\bf z}$}. 
The presence of active forcing linear in $\nabla c$, impossible in the isotropic theory \cite{tiribocchi2015active}, is the central feature of \cref{veq}. \cref{ceq,veq} constitute \emph{Uniaxial Active Model H}. For the remainder of this paper, we will disregard inertia by setting the left-hand side of \cref{veq} to zero, a valid approximation for describing microbial suspensions. We will then see that the velocity field ${\bf u}$ qualitatively changes the concentration dynamics through \eqref{newcurrent} and \eqref{newstress}.  

We begin by neglecting advection ${\bf u} \cdot \nabla c$ in \cref{ceq} and studying the consequences of FIM for the 
effective linearized dynamics of $c$. Defining spatial Fourier components $c_{\bf k}(t) = \int_{\bf r} c({\bf r},t) e^{-i {\bf k} \cdot {\bf r}}$, and similarly for ${\bf u}$, \cref{veq} and incompressibility imply  
\MOD{\begin{align}
    \label{eq:u} 
    {\bf u}_{\bf k}= {\bf B}_{\bf k}c_{\bf k} \equiv - iW \frac{\boldsymbol{\Pi}_{\bf k} }{\eta k^2} \cdot \hat{\bf z} \, k_z \, c_{\bf k} 
    \, \, \mbox{, where} \,\boldsymbol{\Pi}_{\bf k} = \bsf{I} - \hat{\bf k} \hat{\bf k} 
\end{align}
projects transverse to ${\bf k}$.} 
Inserting the result in the Fourier-transformed \cref{ceq} 
gives 
$\partial_t{c_k} = -D(\theta)k^2 c_k$ where $\theta$ is the angle between the wavevector ${\bf k}$ and the $\hat{\bf z}$ axis, with  
\begin{align}
    \label{D(theta)}
    D(\theta) = \left(D_z  -  \frac{bW}{4\eta} \sin^2 2 \theta \right) \cos^2 \theta&&\nonumber \\  + \left(D_\perp -  \frac{aW}{4\eta} \sin^2 2 \theta\right) \sin^2 \theta&&. 
\end{align} 
\SR{The inclusion of} FIM \eqref{newcurrent} stands vindicated as its effect enters \cref{D(theta)} at the same order in wavenumber as the bare diffusive dynamics. 


\MOD{As we remarked above, $a,b$ exist in the ``passive'' hydrodynamic theory \cite{aubert1980macromolecules,callan2011hydrodynamics}. However, the physics of Stokesian swimmers in addition leads to an \textit{active} contribution to $a$ and $b$ which we now examine. To fix ideas let us consider extensile, that is, pusher, swimmers with axes parallel to $\hat{\bf z}$, so that $W>0$, which we can view as contractile (puller) along $\hat{\bf x}$ in \cref{Flow}a and extensile (pusher) along $\hat{\bf z}$ in \cref{Flow}b . 
Given that Stokes drag is proportional to size, the deformation created by the flow (the red arrows) displaces the centre of drag to the right in \cref{Flow}a and upward in \cref{Flow}b. The arguments of \cite{Baskaran15567} then imply motion to the right in \cref{Flow}a and downward in \cref{Flow}b, which implies $a,b>0$ in terms of the  recast currents ${\bf J}_{\perp}^u = a \nabla_{\perp} \partial_z u_z$, $J_z^u = - b \partial_z \nabla_{\perp} \cdot \mathbf{u}_{\perp}$ discussed in and after \eqref{ceq}, \eqref{veq}.}
Similarly, $W < 0$ for $a,b <0$, and so the active parts of $a,b$ obey $aW>0, bW>0$. Thus the active piece of FIM, via \cref{D(theta)}, always \textit{reduces} the diffusivity. For large enough $aW$ or $bW$ the diffusivity will first hit zero for a cone of directions determined by \cref{D(theta)}. Beyond this threshold $D(\theta) < 0$ for a range of $\theta$ -- the anti-diffusion alluded to in the title -- i.e., a linear instability to small-wavenumber concentration fluctuations. Active hydrodynamics thus leads to phase separation without attractive interactions, with structural anisotropy playing the crucial role of orienting the persistent large-scale flows of the swimmers. The condensation can take place in an infinite system, with no need to invoke boundaries \cite{thutupalli2018FlowinducedPhase}. This discussion of course presumes an appreciable magnitude for $a$ and $b$. 
\SR{The migration velocity estimate in Ref. \cite{aubert1980macromolecules} would imply \MOD{passive contributions with magnitude} $a\sim-b\sim c_0R^2$, which means $a$ and $b$ respectively promote instability for extensile and contractile systems.}
In End Matter, we argue \SR{that the additional active contribution discussed above should be of order $(W \tau /\eta R)c_0 $ where $\tau$ is the structural relaxation time of the suspension and $R$ the particle size. We show that the combined relative shift in diffusivities should be of order $\phi Pe[1 + \mathcal{O}(Pe)]$, where $\phi$ is the volume fraction and $Pe = W/\eta D_0 R$ a P\'{e}clet number, which we estimate to be of order unity for plausible parameter values.} See End Matter also for a discussion on possible effects of the hydrodynamic interaction at the putative critical point of this flow-induced condensation. 


We focus now on the homogeneous suspension, for which it suffices to study \eqref{ceq} and  \eqref{veq} with $a, b = 0$, and examine the effects of advection. We will proceed as usual \cite{hohenberg1977theory,Mazenko_noneq} by first establishing the scaling properties of the linear theory, and then testing the stability of the resulting Gaussian fixed point with respect to inclusion of the advective nonlinearity ${\bf u} \cdot \nabla c$. Rescaling ${\bf r} \to b{\bf r}$, \, $t\rightarrow b^z t$, \, $c \to b^{\chi}c$, \, ${\bf u} \to b^{\Xi}{\bf u}, \, \mathcal{P} \to b^{\psi} \mathcal{P}$, 
we find the Gaussian fixed point $z=2$, $\chi = -d/2$ is stable against the advective nonlinearity for $d>4$. 
For $d<4$ the large-scale long-time properties of the homogeneous phase of an anisotropic active suspension are governed by a nontrivial stable fixed point. 
We do not implement a dynamic renormalization-group to obtain the scaling properties for $d<4$. As Galilean invariance ensures the absence of fluctuation corrections to the vertex of the advective nonlinearity, it suffices to carry out a one-loop self-consistent calculation for the correlation function and propagator \cite{Bhattacharjee_1998,bhattacharjee2007non}. 

We define the propagator 
$G({\bf r},t) \equiv \frac{\delta \langle c({\bf r},t)\rangle}{\delta h({\bf 0},0)}\bigg\rvert_{h=0}$ for $t>0$ and $0$ for $t<0$, where $h({\bf r},t)$ is a source field added to the right-hand side of \cref{ceq}, and the correlator $C({\bf r},t) = \langle c({\bf 0},0)c({\bf r},t)\rangle$, 
where the angle brackets are an average over the noise ${\bf f}$ in \cref{ceq}. We work with the space-time Fourier transforms $G_{{\bf k}\omega} \equiv \int_{{\bf r},t}G({\bf r}, t)e^{-i {\bf k} \cdot {\bf r} + i \omega t} \equiv \int_{t}G_{{\bf k}}(t) e^{i \omega t}$, and similarly $C_{{\bf k}\omega}$ and $C_{{\bf k}}(t)$.
The one-loop self-consistent method for evaluating $G$ and $C$ \cite{Bhattacharjee_1998,bhattacharjee2007non} consists in solving \cref{ceq} iteratively by standard methods (see, e.g., \cite{Mazenko_noneq,MaMaz_PhysRevB.11.4077}), giving 
\be
    \label{selfeq} 
    G^{-1}_{{\bf k}\omega} = G^{-1}_{0{\bf k}\omega} - \Sigma_{{\bf k}\omega}
\ee 
where $G_{0{\bf k}\omega} = (-i \omega + D_{\perp} k_{\perp}^2 + D_z k_z^2)^{-1}$ is the bare propagator,  
\be 
    \label{selfenergy} 
    \Sigma_{{\bf k}\omega} =-\lambda^2\int \frac{d^dq}{(2 \pi)^d} dt |V_{\bf k q}|^2 C_{{\bf k} - {\bf q}}(t) G_{{\bf q}}(t) e^{i\omega t}
\ee 
the self-energy, and  
\be
    \label{vertex} 
    V_{\bf kq} = \frac{W}{2\eta}{\bf k} \cdot \left(\boldsymbol{\Pi}_{\bf q}\frac{q_z}{q^2}+ \boldsymbol{\Pi}_{{\bf k} - {\bf q}}\frac{k_z - q_z}{|{\bf k} - {\bf q}|^2}\right) \cdot \hat{\bf z} 
\ee 
the symmetrized vertex. The correlator is expressed in terms of $G$ and the renormalized noise strength $N_{{\bf k} \omega}$ as 
\be 
    \label{correlator}
    C_{{\bf k}\omega} = |G_{{\bf k}\omega}|^2 N_{{\bf k}\omega},
\ee
where 
\be
    \label{noiseren} 
    \!\!N_{{\bf k}\omega} = N_{\perp} k_{\perp}^2 + N_z k_z^2 + \int \frac{d^dq}{(2\pi)^d} dt e^{i \omega t} |V_{\bf k q}|^2 C_{{\bf k} - {\bf q}}(t) C_{{\bf q}}(t).
\ee
Self-consistency is enforced by the use of the full, not bare, propagator and correlator in the integrals in \cref{selfenergy,noiseren}. To extract the scaling exponents $\chi,z$ it suffices to make the dynamical scaling ansatz $G_{{\bf k}}(t) = f(\Gamma k^z t), \,  C_{{\bf k}}(t) = A k^{-(d + 2 \chi)} g({\Gamma k^z}t)$,
with coefficients $A$ and $\Gamma$ for the static correlator and the relaxation rate. We approximate the scaling functions $f(x)$ and $g(x)$ by $e^{-x}$, i.e., a single relaxation time for each wavenumber or equivalently a Lorentzian line-shape in the frequency domain, which is acceptable if the aim is only to identify scaling exponents. For $d<4$, relaxation at small $k$ is dominated by the self-energy, so we identify $\Sigma_{{\bf k} \omega} = \Gamma k^z$ and $N_{{\bf k}\omega}$ by the integral in \cref{noiseren}. Power-counting on \cref{selfenergy} can readily be seen 
to yield one condition on the two exponents, namely, 
\be
    \label{galinv}
    \chi + z = 0. 
\ee
\cref{galinv}, which we obtained here by assuming the dominance of the fluctuation correction in the self-consistent calculation, is precisely the result that would emerge if we invoked Galilean invariance to keep the coefficient of ${\bf u} \cdot \nabla c$ in \cref{ceq} fixed under rescaling. Scaling all wavevectors by $k$ itself and inserting the scaling ansatz 
allows us to rewrite \cref{selfenergy,noiseren} as 
\begin{align}
    \label{SC1}
    \frac{2 \Gamma^2}{A} &=  \int \frac{d^dq}{(2 \pi)^d} |V_{\hat{\bf k} \hat{\bf q}}|^2 \frac{1}{q^2} \frac{1}{|\hat{\bf k} - {\bf q}|^{d + 2 \chi}} \frac{1}{q^z} \nonumber \\
    &\quad \asymp \tilde{K}_d \frac{1 - \Lambda^{-(2 + 2 \chi + z)}}{2 + 2 \chi + z},
\end{align}
and 
\begin{align}
    \label{SC2}
    \frac{2 \Gamma^2}{A} &=  \int \frac{d^dq}{(2 \pi)^d} |V_{\hat{\bf k} \hat{\bf q}}|^2 \frac{1}{q^2}  \frac{1}{|\hat{\bf k} - {\bf q}|^{d + 2 \chi}} \frac{1}{q^{d+2\chi}} \frac{2}{|\hat{\bf k} - {\bf q}|^z + q^z} \nonumber \\
    &\quad \asymp \tilde{K}_d \frac{1 - \Lambda^{-(2 + 3 \chi + d)}}{2 + 3 \chi + d},
\end{align}
where we retain the symbol ${\bf q}$ for the rescaled internal wavevector and have scaled the powers of $q$ out of the vertex. In \cref{SC1,SC2}, $\Lambda$ is the ultraviolet cutoff, which we can take to $\infty$ for $d<4$, $\tilde{K}_d$ is an angular integral, and the asymptotic equality ($\asymp$) in each case is the large-$q$ contribution which dominates for $d \to 4$ (and turns into the logarithmic ultraviolet divergence for $d=4$) \cite{Bhattacharjee_1998,bhattacharjee2007non}. Equating these for $d = 4 - \epsilon$, we see that 
\be
    \label{chi+2d}
    d + 2\chi = 0, 
\ee
meaning that within our isotropic scaling ansatz the \textit{spatial} rescaling properties of $c$ are always those of the Gaussian theory. The dynamical scaling properties are however nontrivial: combining \cref{chi+2d} with \cref{galinv} implies a superdiffusive dynamic exponent   \be
    \label{dynexp}
    z = 2 - \epsilon/2,  
\ee 
i.e., time $\sim$ length$^{3/2}$ in $d=3$, or mean-square displacement $\sim t^{4/3}$.  

Two arguments, one intuitive, the other formal, help us understand the self-consistent scaling results and their limitations. 
\MOD{Along lines similar to \cite{Caflisch-Luke}, consider a suspension of $N$ force dipoles of identical strength $W$, uniaxially aligned in a fluid with viscosity $\eta$, distributed randomly and uniformly within a domain of linear dimension $R$.
Their mean number density is $c_0 \sim N /R^d$, where 
$d$ is the space dimension.
Each such dipole gives rise to a flow field of magnitude  $\tilde{u} \sim R^{-(d-1)}W/\eta$. 
Assuming 
independent concentration fluctuations, we can add their individual contributions to get the $R$-dependent part $[\delta u(R)]^2 \sim N \tilde{u}^2 \sim c_0(W/\eta)^2 R^{-d+2}$ of the velocity variance (see End Matter for detailed calculation).}
The corresponding timescale is $\tau = R/\delta u(R) \sim (\eta/W c_0^{1/2}) R^{d/2}$, which trumps diffusion if $d < 4$, reproducing both the dynamic exponent and the upper critical dimension that we reported above. Of course this argument can't tell us that it was correct to treat the swimmers as independent. Our self-consistent calculation yielded both $z=d/2$ and $\chi = -d/2$, the latter implying the absence of positional correlations at large length scales, unlike in sedimentation, where the hydrodynamic interaction drastically alters \cite{levine1998screened,sedreview,kumar2011nonequilibrium,koch1991screening} the independent-particle calculation \cite{Caflisch-Luke}.

More formally, if we set $a=b=0$ in \cref{ceq}, and choose the \textit{bare} noise and diffusion to obey $N_z/N_{\perp} = D_z/D_{\perp}$, then the steady-state solution of the Fokker-Planck equation for the probability distribution functional for the field $c$ in the \textit{linear} theory (i.e., ignoring advection) is $P_s[c] \propto \exp [-\mbox{const} \int_{\bf x} (\delta c)^2]$ so that the static correlator $C_{\bf k}(0)$ is a constant, i.e., independent of ${\bf k}$. 
Does this result survive the inclusion of advection? It does, because \MOD{as shown in detail in End Matter} the functional divergence $\int_{\bf x}{\delta}J/{\delta c({\bf x})}$ of the probability current $J = {\bf u} \cdot \nabla c P_s[c]$ that the advective term ${\bf u} \cdot \nabla c$ induces, given the distribution $P_s$, is identically zero.
Thus $P_s[c] \propto \exp [-\mbox{const} \int_{\bf x} (\delta c)^2]$ is an exact steady-state solution to the 
Fokker-Planck equation even when the advective nonlinearity is included. Then $\chi = -d/2$ and therefore, for $d \le 4$, as Galilean invariance implies $\chi + z = 0$, the dynamic exponent $z=d/2$. If $N_z/N_{\perp} \neq D_z/D_{\perp}$, these arguments do not apply, and a more complicated scenario emerges which we discuss briefly in End Matter.
\begin{figure}
    \hspace{-0.7cm}
    \includegraphics[width=\columnwidth]{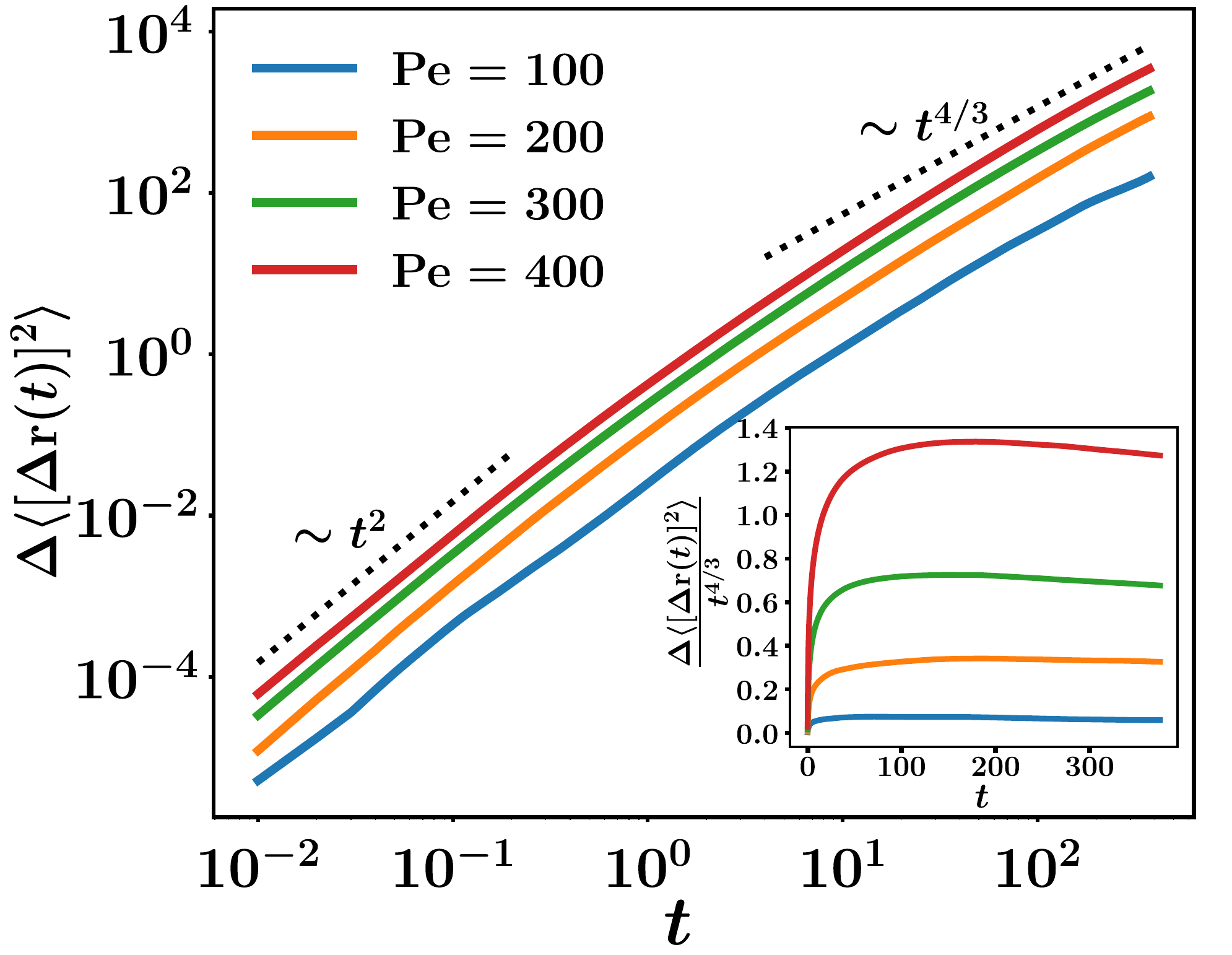} 
    \caption{Mean-Square-Displacement(MSD) of the particles, after substracting diffusive contributions, 
    in log-log scale, 
    (\textit{inset})
    in linear-linear scale after dividing by corresponding superdiffusive scaling, for different P\'eclet numbers. }
    \label{NumericsPlots}
\end{figure} 

We now test our field-theoretic predictions against a numerical simulation of Brownian point force dipoles of strength $W$, situated at positions $\{\mathbf{r}_\alpha(t), \, \alpha = 1, ... N\}$, aligned in the $\hat{\bf z}$ direction, suspended in a Stokesian fluid with viscosity $\eta$ and velocity field $\mathbf{u}(\mathbf{r})$, and obey the advected Brownian equation of motion 
\begin{align} 
\label{advBrown}
\dot{\mathbf{r}}_\alpha = \mathbf{u}(\mathbf{r}_\alpha) + \sqrt{2D} \; \boldsymbol{\zeta}_\alpha(t), 
\end{align}
where $D$ is the diffusivity and $\boldsymbol{\zeta}(t)$ is isotropic unit Gaussian white noise. 
We work with a 3D domain with periodic boundary conditions. Scaling lengths by the simulation grid size $\ell$ and times by $\ell^2/D$, the non-dimensionalized Eulerian velocity field due to the force dipoles is $\mathbf{u}(\mathbf{r}) = \mathrm{Pe} \sum_{\mathbf{k}}\sum_{\alpha} {ie^{i\mathbf{k}\cdot(\mathbf{r-r_{\alpha}})}}\big(k_z^{2}\mathbf{k_{\perp}} - k_{\perp}^2 k_z\hat{\bf{z}}\big)/{k^4}$, with P\'eclet number $\mathrm{Pe} = W/\eta D \ell$. 
The velocity field $\mathbf{u}$ evaluated on the simulation grid is interpolated to the $\{\mathbf{r}_\alpha(t)\}$ using a 4-point immersed boundary kernel \cite{peskin2002immersed}.

We consider a cubic box of size $200^3$, 
seed 4000 particles uniformly throughout the system and run the simulation for $10^6$ iterations with a time step of $\delta t = 5 \times 10^{-4}$, using the Euler-Maruyama algorithm 
(see Supplemental Material \cite{supplement} for further details).

\nocite{dunweg1993MolecularDynamics, chandler1988introduction, hansen2013theory, allen2017computer}

\cref{NumericsPlots} 
displays the excess Mean-Square Displacement (MSD) with respect to ordinary diffusion (hereafter the ``filtered MSD''), $\Delta\langle [\Delta \mathbf{r}(t)]^2\rangle \equiv \langle |\mathbf{r}(t)-\mathbf{r}(0)|^{2}\rangle - 6t$ in the statistical steady state. We see an important feature apart from the asymptotic long-time superdiffusion, namely, an early-time ballistic regime, which we now show is also predicted by our theory.   
The filtered MSD of the particles is determined entirely by the velocity field generated by the force dipoles: 
$\Delta\langle [\Delta \mathbf{r}(t) ]^2 \rangle = \int_{s=0}^t ds \int_{s'=0}^t ds' \; \big\langle \mathbf{u}(\mathbf{r}_{\alpha}(s), s) \cdot \mathbf{u}(\mathbf{r}_{\alpha}(s'), s')$. Eq. \eqref{veq} tells us that the velocity correlator in the integrand has a well-defined $s\to s'$ limit which, for $t \to 0$, constitutes the coefficient of $t^2$, the leading short-time behavior of the filtered MSD \cite{supplement}.
\cref{NumericsPlots} (\textit{inset})
shows the filtered MSD divided by the proposed asymptotic superdiffusive contribution to the MSD (i.e., $L^2 \sim \tau^{\frac{4}{3}}$ for 3D), revealing a plateau that persists over a long time, confirming the prediction \eqref{dynexp}. 
Note that the noise-averaged particle-phase velocity field due to \cref{advBrown} is divergenceless and therefore cannot generate the physics of FIM \eqref{newcurrent} and the diffusive instability \eqref{D(theta)}. The latter would require motion across streamlines of ${\bf u}$, as would arise in the presence of interparticle interactions via a pair potential. 

To summarise: we have shown that the large-scale, long-time properties of a suspension of active particles in a viscous fluid with a built-in preferred axis lie in universality classes distinct from those of an isotropic active suspension with the same slow variables. Unlike in \cite{tiribocchi2015active}, activity enters even in the linearized dynamics of the concentration and hydrodynamic velocity. In the homogeneous phase, we predict superdiffusive relaxation of the concentration, with dynamic exponent $z=d/2$ in dimension $d<4$, i.e., time $\propto$ (length)$^{3/2}$ in 3D. Our 3D numerical experiments with aligned Brownian force dipoles in a viscous fluid show a crossover from an early-time ballistic to the asymptotic superdiffusive behavior, confirming our proposed scaling. We have \SR{shown that the interplay of active stresses and a particle current induced by inhomogeneous gradients in the suspension velocity field} 
drives phase separation through hydrodynamic flow. \SR{In End Matter, we argue that the relative shift in diffusivities is of order the P\'{e}clet number associated with activity, which for reasonable parameter values is found to be of order unity.} The underlying mechanism is fundamentally multi-particle in nature and thus inaccessible to the treatment of \cite{Baskaran15567}. By contrast with the flow-induced phase separation of \cite{thutupalli2018FlowinducedPhase}, it does not rely on sample boundaries.
Possible physical realizations on which to test our theory include microbes dispersed homogeneously in a nematic \cite{Zhou1265} stiff enough that its orientational instability due to active stress \cite{simha2002hydrodynamic} is banished to scales beyond the system size—, or subjected to polarized \cite{yang2021controlling} or directional \cite{eisenmann2025PureHydrodynamic,eisenmann2023phototaxis,choudhary2019reentrant} light fields, or colloidal suspensions in a uniform electric field \cite{zorkot2018current,mahdisoltani2021long,golestanian2026AnomalousDiffusion}, in which concentration fluctuations are force dipoles. Similar physics, though accompanied by long-ranged magnetic effects, should arise for magnetic colloids in an oscillating magnetic field~\cite{kumaran2022effect}.

\emph{Acknowledgments—} 
We thank J K Bhattacharjee, A Callan-Jones, R Golestanian, F J\"{u}licher, A Maitra and P Popli for valuable discussions. SKS acknowledges computational support from Supercomputer Education and Research Centre, IISc. SR acknowledges support from a JC Bose Fellowship and a National Science Chair of the ANRF, India, and the Isaac Newton Institute for Mathematical Sciences for support and hospitality during the programmes ``New statistical physics in living matter: non equilibrium states under adaptive control'' and ``Anti-diffusive dynamics: from sub-cellular to astrophysical scales'', supported by EPSRC Grant No. EP/R014604/1, and for a Rothschild Distinguished Visiting Fellowship. SR also acknowledges the support and hospitality of the workshop ``From Soft Matter to Biophysics'', Les Houches (2023). KVK acknowledges support of the Department of Atomic Energy, Government of India, under Project No.~RTI4019.


\bibliography{bibli} 

\clearpage 
\section*{ {\bf End Matter}}

\subsection{Flow-induced \SR{migration}} \label{FIM_end}
\subsubsection{General structure} \label{FIM_structure}
We now discuss ${\bf J}^u$ [\eqref{newcurrent}] and its dependence on particle-scale properties. The options for its form are limited: the absence of a slow variable associated with orientational distortions rules out currents familiar from flocking models \cite{toner1998flocks} and active nematics \cite{SRAditiToner2003nematic}; power-counting \cite{supplement} shows that currents nonlinear in $\nabla c$ \cite{activeB_NComm,tiribocchi2015active,activeB+bubbly_PRX} are irrelevant to the scaling properties of interest here; Galilean invariance tells us that the only current that a \textit{uniform} velocity field can produce is $c {\bf u}$. 
\SR{A velocity profile with nonzero \textit{curvature}, such as a local Poiseuille flow, however, simultaneously breaks time-reversal and defines a vectorial asymmetry in a Galilean-invariant manner (see the supplement of \cite{maitra2014activating} and \cite{chajwa2023active}). Thus a particle current proportional to the curvature cannot be ruled out on symmetry grounds \cite{Curie_Principle}, so we must allow for its existence. 
In other words, there should in general be a contribution to ${\bf J}^u$ of the form 
\begin{align}
    \label{newcurrentold}
    J_i^u = A_{ijkl} \nabla_j \nabla_k u_l 
\end{align} 
where $A_{ijkl}$ 
are the components of a general 4th rank tensor $\bsf{A}$, symmetric in $jk$, constructed only from the unit tensor and $\hat{\bf z}$. 
It is similar in spirit to a current \cite{kumar2011nonequilibrium,Goldfriend2017} associated with screening in sedimenting suspensions \cite{levine1998screened,sedreview} and to the current proportional to orientational curvature in active nematics \cite{SRAditiToner2003nematic}. A particle current of the form \eqref{newcurrentold} but with isotropic $A_{ijkl}$ is permitted for dynamics at equilibrium, with, in our notation, $a,b \sim c_0R^2$ where $R$ is a typical particle size \cite{callan2011hydrodynamics,aubert1980macromolecules}. However, anisotropy is crucial for the effect we discuss: in the \textit{isotropic} limit ${\bf J}^u \sim c \nabla^2 {\bf u} + \ldots$, in which case incompressibility implies $\nabla \cdot {\bf J}^u \sim \nabla c \cdot \nabla^2 {\bf u}$, with no linear contribution.}

\subsubsection{\texorpdfstring{Estimating magnitudes}{a, b values estimation}} We have already remarked that the FIM coefficients $a$ and $b$ arise within passive, that is, equilibrium, dynamics \cite{callan2011hydrodynamics,aubert1980macromolecules}, with a magnitude $\sim c_0R^2$. We now show below that they also receive contributions from active processes. 
%
Consider a dense suspension with positional correlations on a scale $R$ which one may take to be the inverse of the location of the peak of the static structure factor, which is of order the particle size, and structural relaxation time $\tau$. A velocity gradient $\nabla {\bf u}$ produces anisotropy of order $\tau \nabla {\bf u}$ (times a numerical factor reflecting the wavevector dependence of the structure factor, which we do not display) \cite{clark1980observation,ronis1984theory,pryde1966liquid}. Inhomogeneity $\nabla \nabla {\bf u}$ in the velocity gradient produces polarity $\sim \tau R \nabla \nabla {\bf u}$ in interparticle correlations. Arguing as in \cite{Baskaran15567}, the active force scale $W/R$ [\cref{newstress}, \cref{veq}] then implies a polar forcing of order $\tau R \nabla \nabla {\bf u} W/R$ which, divided by viscous drag $\sim \eta R^{d-2}$, yields a velocity $\tau  \nabla \nabla {\bf u} W/\eta R^{d-2}$. This implies a current $(c_0 \tau  W/ \eta R^{d-2}) \nabla \nabla {\bf u}$ where $c_0$ is the typical concentration, and thus the estimate $a,b \sim c_0 \tau  W/ \eta R^{d-2} = c_0 \tau W /\eta R$ for $d=3$. Adding this active piece to the passive contribution mentioned above gives $a,b \sim c_0 R^2(1 + \tau W /\eta R^3)$. \SR{Note: in the language of \cite{de2013non} the current \eqref{newcurrent} represents the ``flux'' ${J}^u$ due to a ``force'' ${\bf u}$. A $W$-dependent input to this term has a further dependence on another ``force'', namely, the chemical driving that maintains the active stress (see, e.g., \cite{Prawar_Dadhichi_2018}), and is thus a contribution at second order in ``forces'', not strictly in the domain of \textit{linear} irreversible thermodynamics \cite{de2013non}.}

\SR{The resulting relative shift in diffusivity in \cref{D(theta)} is $a W /\eta D_0 \sim \phi(Pe + Pe^2)$, where the second order term is due to the active contribution to $a,b$, and we have defined an active P\'{e}clet number $Pe \equiv W/\eta D_0 R$ and volume fraction $\phi = c_0 R^3$, and assumed $\tau \sim R^2/D_0$ times an increasing function of volume fraction which we take to be of order unity.
Using thermal motion to estimate $D_0$ for active particles is unrealistic. Swim speed $v_0 \simeq 10 \, \mu\mathrm{m}/$s and a run time of $\tau_R = 1 \, \mathrm{s}$ gives $D_0 \sim v_0^2 \tau_R \sim 10^{-6}$ cm$^2$ s$^{-1}$. $W \sim 1$ pN $\mu$m \cite{dunstan2012two}, $R\sim 1 \, \mu$m gives the encouraging estimate $Pe \sim 1$.
}  


\subsection{Condensation by FIM under quasi-2D confinement} \label{sub:FIM_confined}
Note that even when the system is confined by walls to a 2D fluid film \cite{maitraNonequilibriumForceCan2018}, the active currents generated by the averaged flow field along the confinement direction still contribute diffusion-like terms to the continuity equation \cref{ceq}. This results in the possibility of condensation through the same mechanism \cite{supplement}. Furthermore, although we assumed instantaneous steady Stokesian dynamics above, introducing viscous retardation for the velocity field -- i.e., considering the unsteady Stokes equation -- still leads to condensation with precisely the same diffusivity shifts as obtained earlier \cite{supplement}. 

\subsection{Condensation by FIM: relevance of the hydrodynamic interaction} \label{sub:hydro_relevance}
Now we return briefly to the effective diffusivity due to FIM. \cref{D(theta)} tells us there is a locus in parameter space at which the diffusivity first hits zero. We do not know if the conventional scenario of 1st-order coexistence, spinodal line and critical point apply to this active system. However, if a point on the zero-diffusivity locus is accessible by tuning a control parameter, it is like a critical point of phase separation. As $D(\theta)$ vanishes for a particular value of $\theta$, this critical point is highly anisotropic, with static structure factor (given by the ratio of the noise strength to the relaxation
rate) diverging only for directions of wavevector on a subspace of dimension $d-1$. The  
simplified case of a structure factor $\propto k^2 / (k_z^2 + \lambda^2 k_{\perp}^4)$, which diverges as $1/k^2$ only for $k_z=0$, so that the subspace is a plane, suggests what might be in store: Integrals over ${\bf k}$ receive appreciable contributions only from a width in $k_z$ of order $\lambda k_{\perp}^2$, so that the integration element $d^dk = dk_z d^{d-1}k_{\perp} \sim \lambda k_{\perp}^2 d^{d-1}k_{\perp} \sim k_{\perp}^d dk_{\perp}$, which is as though the system were in $d+1$ dimensions. At this mean-field critical point for a conserved order parameter, $z = 4$ and the exponent $\chi = (1 - d)/2$, corresponding to a structure factor diverging as $1/k^2$ in dimension $d+1$. We see then that the rescaling factor $\chi + z$ for the advective vertex in \cref{ceq} is $(9-d)/2$, i.e., the vertex is i.e., effects associated with the hydrodynamic interaction are relevant for all dimensions $d < 9$, overwhelmingly so for $d=3$. 
\MOD{
\subsection{Comparison and contrast to sedimentation
} 
In a spirit similar to Caflisch and Luke's treatment of velocity fluctuations in sedimentation \cite{Caflisch-Luke}, consider a fluid of viscosity $\eta$ containing a suspension of $N$ force dipoles of strength $W$ and excluded volume $a^d$, uniaxially aligned along $\hat{\mathbf{z}}$ and distributed independently and uniformly within a domain $\Omega_R$ with boundary $\partial\Omega_R$ of linear dimension $R$.
Their mean number density $c_0 \sim N /R^d$, where 
$d$ is the spatial dimension.
The total Stokes flow $\mathbf{u}$ inside $\Omega_R$ is $\mathbf{u}(\mathbf{r}) = (W/\eta)\sum_{n=1}^{N}\tilde{\mathbf{u}}(\mathbf{r},\mathbf{r}_n)$, where $\tilde{\mathbf{u}}(\mathbf{r},\mathbf{r}_n)$ accounts for the flow at position $\mathbf{r}$ generated by a dipole located at $\mathbf{r}_n$, and is just the $z$ derivative of a unit Stokeslet. For $\mathbf{x},\mathbf{y}\in\Omega_R$, 
\begin{equation}
\label{singleStokes}
    \nabla^2\tilde{\mathbf{u}} + \nabla\mathcal{P}  = \hat{\bf z} \partial_z \delta({\bf x} - {\bf y}), \quad \nabla\cdot\tilde{\mathbf{u}} = 0\; \text{in}\; \Omega_R,
\end{equation}
and $\tilde{\mathbf{u}} = 0\; \text{on} \; \partial\Omega_R$, with $\mathcal{P}$ the corresponding pressure. 
For $\mathbf{r}_n$ uniformly and independently distributed over $\Omega_R$, the average of $\tilde{\mathbf{u}}(\mathbf{r},\mathbf{r}_n)$ vanishes as it is the $z$ derivative of that in \cite{Caflisch-Luke}. What interests us is the variance 
\begin{equation} 
\label{variance}
        \langle|\mathbf{u}|^2\rangle 
        = \left(\frac{W}{\eta}\right)^2\sum_{n=1}^N\langle|\tilde{\mathbf{u}}(\mathbf{r},\mathbf{r}_n)|^2 \rangle \sim \left(\frac{W}{\eta}\right)^2 c_0 \int_{\bf r} r^{2-2d}\, 
\end{equation}
in $d$ space dimensions. Keeping in mind that the integration range is $a < r < R$, we see that $\langle|\mathbf{u}|^2\rangle \sim (W/\eta)^2 c_0 (a^{2-d} - R^{2-d})$, so that the approach of velocity fluctuations to their $R=\infty$ value is $[\delta u(R)]^2 \equiv \langle|\mathbf{u}|^2\rangle_{R = \infty} -  \langle|\mathbf{u}|^2\rangle_{R} \sim (W/\eta)^2 c_0 R^{2-d}$, $\sim 1/R$ in $d=3$. Thus, unlike in sedimentation \cite{sedreview}, velocity fluctuations are finite in the infinite-size limit. We expect, however, that their effect on concentration dynamics arises from the scale-dependent correlations contained in $\delta u(R)$, through the characteristic timescale $R/\delta u(R) \sim R^{d/2}$. Reassuringly, this argument reproduces the result of our self-consistent calculation. 
}
\subsection{\MOD{Functional Fokker-Planck Equation}} \label{sub:FP_int} 
\MOD{
The stochastic PDE \eqref{ceq} for the concentration-field $c$ is equivalent to a Fokker–Planck equation for the probability functional $P[c]$, $\partial_t P[c] = - \int d^d r \, \delta J / \delta c({\bf r})$, where $J$ is the probability current in $c$ space generated by the dynamics \eqref{ceq}.
For any distribution $P[c]$, the advective term in \cref{ceq} contributes a drift $-P[c] {\bf u} \cdot \nabla c $ to $J$, with ${\bf u}$ determined by $c$ via \cref{eq:u}. 
The functional divergence of this piece of the current is}  
\begin{align}
    \label{divprobcurr}
    \int_{\bf x} \frac{\delta}{\delta c({\bf x})} ( P[c]\,{\bf u} \cdot \nabla c)&=\int_{\bf x}\left[ {\bf B}({\bf 0}) \cdot \nabla c + {\bf u} \cdot \nabla \delta({\bf 0}) \right] P[c] \nonumber \\
    &\quad + \int_{\bf x}  {\bf u} \cdot \nabla c \frac{\delta P}{\delta c ({\bf x})}, 
\end{align}
\MOD{where we have expressed ${\bf u}$ in terms of $c$ through the kernel ${\bf B}$ in \cref{eq:u}. On the right-hand side of \cref{divprobcurr}, the terms in square brackets vanish: ${\bf B}({\bf 0}) = \int_{\bf k} {\bf B}_{\bf k} = {\bf 0}$ because ${\bf B}_{\bf k}$ is odd in ${\bf k}$ and the gradient of the Dirac delta vanishes at zero argument. The remaining term vanishes whenever $P$ is purely local, i.e., if $\log P = \int_{\bf x} g(c({\bf x}))$ for any function $g$ of $c$ alone and not its gradients, as it can be rewritten, integrating by parts and using incompressibility, as $P_s[c] \int_{\bf x} \nabla \cdot ({\bf u}  g)$ which becomes a boundary term. This property holds for the stationary solution $P_s[c] \propto \exp [-\mbox{const} \int_{\bf x} (\delta c)^2]$ in the absence of the advective term, or more generally in a microscopically detailed stochastic description \cite{Dean1996}. Thus $P_s$ remains a stationary solution to the FPE even when advection is included, and the advective nonlinearity doesn't affect the static scaling of the linear theory.}

If on the other hand $N_z/N_{\perp} \neq D_z/D_{\perp}$, the static structure factor in the linear theory will depend on the direction (though not the magnitude) of the wavevector \cite{PhysRevLett.64.1927}. A distribution depending only on $c$ and not its gradients will not be a stationary solution to the Fokker-Planck equation even for the linear theory, and the line of reasoning above will not go through.
Moreover, fluctuation corrections due to the advective nonlinearity will generate the $a$ and $b$ terms in \cref{ceq} even if these were absent in the bare theory, without a corresponding noise correction \cite{supplement}, in a manner precisely analogous to the generation of screening terms in sedimentation \cite{levine1998screened,kumar2011nonequilibrium},
with coefficients proportional to $N_\perp/D_\perp- N_z/D_z$, highlighting their nonequilibrium character.
Note, however, that even in this case the anisotropy in the linear theory is weak: for all directions of wavevector ${\bf k}$ the relaxation rate is $O(k^2)$ and the structure factor is $O(1)$. Anisotropic scaling, if any, can emerge only from the nonlinear theory. It is possible then that the isotropic scaling solution has a domain of attraction larger than that defined by the restriction $N_z/N_{\perp} = D_z/D_{\perp}$. Settling this issue, however, requires a self-consistent or renormalization-group treatment with anisotropic scaling, which we defer to later work.

\let\bibliography\relax

\pagebreak
\widetext
\newpage


\section*{Supplementary material}

# Supplemental Material: Superdiffusion and antidiffusion in an aligned active suspension

## I. POWER COUNTING NONLINEAR TERMS

In this section, we show that contributions to the stress nonlinear in $c$ and its gradients, and noise in the Stokes equation, are irrelevant in the Renormalization Group (RG) sense. Using notation from the main text, writing the Navier-Stokes equation for the suspension velocity field by including a force density whose form is that of the divergence of an interfacial stress higher order in gradients and concentration field $c$, and a conserving additive noise, we have

$$\rho(\partial_t + \mathbf{u} \cdot \nabla)\mathbf{u} = \eta\nabla^2\mathbf{u} - \nabla\mathcal{P} - W\hat{\mathbf{z}}\partial_z c + \nabla \cdot (\nabla c\nabla c) + \nabla \cdot \mathbf{f}, \tag{S1}$$

where $\mathbf{f}$ is the Gaussian white noise. If we rescale $\mathbf{r} \to b\mathbf{r}$, $t \to b^z t$, $c \to b^\chi c$, $\mathbf{u} \to b^\Xi\mathbf{u}$, $\mathcal{P} \to b^\psi\mathcal{P}$, we get

$$\rho(\partial_t + b^{z+\Xi-1}\mathbf{u} \cdot \nabla)\mathbf{u} = b^{z-2}\,\eta\,\nabla^2\mathbf{u} - b^{z-\Xi+\psi-1}\nabla\mathcal{P} - b^{z-\Xi+\chi-1}\,W\hat{\mathbf{z}}\,\partial_z c$$
$$+ b^{z-\Xi+2\chi-3}\nabla \cdot (\nabla c\nabla c) + b^{\frac{z-d}{2}-\Xi-1}\nabla \cdot \mathbf{f}, \tag{S2}$$

and the continuity equation for $c$ becomes

$$(\partial_t + b^{z+\Xi-1}\mathbf{u} \cdot \nabla)c = b^{z-2}\left(D_z\partial_z^2 + D_\perp\nabla_\perp^2\right)c - b^{\Xi+z-\chi-3}\left(a\nabla_\perp^2 + b\partial_z^2\right)\partial_z u_z + b^{\frac{z-d}{2}-\chi-1}\nabla \cdot \mathbf{f}. \tag{S3}$$

In the Stokesian limit (droping the left-hand side of Eq. (S2)), the choice $z = 2$, $\chi = -d/2$ and $\psi = \chi = \Xi - 1$ ensures that the coefficients of all terms other than the advective nonlinearity in Eq. (S3) remain unchanged under rescaling, and the noise correlator in Eq. (S3) takes the same form in the rescaled variables as it did in the original variables. We have thus found the Gaussian fixed point for our problem with critical dimension for advective nonlinearity $d_c = 4$. At this fixed point, we see that the exponents of the additional higher-order term and noise in Eq. (S1), namely $z - \Xi + 2\chi - 3$, and $\frac{z-d}{2} - \Xi - 1$, are negative and therefore irrelevant under RG. Each additional factor of $c$ gives an additional $b^\chi$, which decreases under rescaling, as $\chi = -d/2$ for our self-consistent solution. Terms with higher powers of concentration, even with only one $\nabla$, are thus irrelevant.

## II. EMERGENCE OF FLOW INDUCED MIGRATION TERMS

In this section, using diagrammatic perturbation theory up to one-loop order [1], we demonstrate that fluctuation corrections arising from the advective nonlinearity generate the *Flow-Induced Migration* (FIM) terms $a$ and $b$, as introduced in the main text, even if these terms are absent in the bare theory.

For concentration field $c(\mathbf{r}, t)$ and fluid velocity field $\mathbf{u}(\mathbf{r}, t)$, the dynamical equations are

$$(\partial_t + \lambda\mathbf{u} \cdot \nabla)c(\mathbf{r}, t) = (D_{0\perp}\nabla_\perp^2 + D_{0z}\nabla_z^2)c(\mathbf{r}, t) + f(\mathbf{r}, t), \tag{S4}$$

and

$$\eta\nabla^2\mathbf{u} = \nabla\mathcal{P} + W\mathbf{Q} \cdot \nabla c, \quad \nabla \cdot \mathbf{u} = 0, \tag{S5}$$

with

$$\langle f(\mathbf{r}, t)f(\mathbf{r}', t')\rangle = (N_{0\perp}\nabla_\perp^2 + N_{0z}\nabla_z^2)\delta^d(\mathbf{r} - \mathbf{r}')\delta(t - t'). \tag{S6}$$

With the following definition of the Fourier transform,

$$c(\vec{\mathbf{k}}) \equiv c(\mathbf{k}, \omega) = \int d^d\mathbf{r} \int dt\, c(\mathbf{r}, t)e^{-i(\mathbf{k}\cdot\mathbf{r}-\omega t)}, \tag{S7}$$

where $\vec{\mathbf{k}}$ is a short notation for $(\mathbf{k}, \omega)$, Eqs. (S4) and (S5) takes the form

$$-i\omega\, c(\vec{\mathbf{k}}) = -i\lambda \int \frac{d^d\mathbf{q}d\Omega}{(2\pi)^{d+1}}\mathbf{u}(\vec{\mathbf{q}}) \cdot (\mathbf{k} - \mathbf{q})c(\vec{\mathbf{k}} - \vec{\mathbf{q}}) - (D_{0\perp}k_\perp^2 + D_{0z}k_z^2)c(\vec{\mathbf{k}}) + f(\vec{\mathbf{k}}), \tag{S8}$$

$$u_l(\vec{\mathbf{k}}) = \frac{-iWP_{il}(\mathbf{k})k_j\mathsf{Q}_{ij}}{\eta k^2}c(\vec{\mathbf{k}}), \tag{S9}$$

with $P_{ij}(\mathbf{k}) = \delta_{ij} - k_ik_j/k^2$ is transverse projector enforcing the incompressibility constraint of the fluid, and the noise term satisfying

$$\langle f(\vec{\mathbf{k}})f(\vec{\mathbf{k}}')\rangle = (N_{0\perp}k_\perp^2 + N_{0z}k_z^2)\delta^d(\vec{\mathbf{k}} + \vec{\mathbf{k}}') \equiv \mathcal{N}_0(\mathbf{k})\delta^d(\vec{\mathbf{k}} + \vec{\mathbf{k}}'). \tag{S10}$$

Using Eq. (S9) in equation Eq. (S8), we get

$$c(\vec{\mathbf{k}}) = G_0(\vec{\mathbf{k}})f(\vec{\mathbf{k}}) - \frac{\lambda W}{\eta}G_0(\vec{\mathbf{k}})\int\frac{d^d\mathbf{q}d\Omega}{(2\pi)^{d+1}}\frac{k_lP_{li}(\mathbf{q})q_j\mathsf{Q}_{ij}}{q^2}c(\vec{\mathbf{q}})c(\vec{\mathbf{k}} - \vec{\mathbf{q}}), \tag{S11}$$

where the bare propagator

$$G_0(\vec{\mathbf{k}}) = [-i\omega + \tau_o^{-1}(\mathbf{k})]^{-1}, \tag{S12}$$

and the bare relaxation time

$$\tau_0(\mathbf{k}) = [D_{0\perp}k_\perp^2 + D_{0z}k_z^2]^{-1}. \tag{S13}$$

We define bare correlator

$$C_0(\vec{\mathbf{k}}) \equiv \mathcal{N}_0(\mathbf{k})G_0(\vec{\mathbf{k}})G_0(-\vec{\mathbf{k}}). \tag{S14}$$

We now write Eq. (S11) as

$$c(\vec{\mathbf{k}}) = G(\vec{\mathbf{k}})f(\vec{\mathbf{k}}), \tag{S15}$$

where $G(\vec{\mathbf{k}})$ is the full propagator. Diagrammatically, we can represent the equation of motion in Fourier space and the corresponding three point and two point vertices as shown in Figs. S1(a-c). With the full propagator given by the Dyson equation

$$G^{-1}(\vec{\mathbf{k}}) = G_0^{-1}(\vec{\mathbf{k}}) + \left(\frac{\lambda W}{\eta}\right)^2\Sigma(\vec{\mathbf{k}}), \tag{S16}$$

which is represented diagrammatically in Fig. S1(d), and the self-energy, $\Sigma(\vec{\mathbf{k}})$, at one-loop order is as shown in Fig. S1(e). This translates to

$$\begin{aligned}
\Sigma(\vec{\mathbf{k}}) = \int\frac{d^d\mathbf{q}d\Omega}{(2\pi)^{d+1}}\frac{k_lP_{li'}(\mathbf{q})q_{j'}\mathsf{Q}_{i'j'}}{q^2}\Bigg[&\frac{(k_m - q_m)P_{mi}(\mathbf{q})(-q_j)\mathsf{Q}_{ij}}{q^2}C_0(\vec{\mathbf{q}})G_0(\vec{\mathbf{k}} - \vec{\mathbf{q}}) \\
+&\frac{k_mP_{mi}(\mathbf{q} - \mathbf{k})(q_j - k_j)\mathsf{Q}_{ij}}{(\mathbf{q} - \mathbf{k})^2}C_0(\vec{\mathbf{k}} - \vec{\mathbf{q}})G_0(\vec{\mathbf{q}}) \\
+&\frac{(k_m - q_m)P_{mi}(\mathbf{k})k_j\mathsf{Q}_{ij}}{k^2}C_0(\vec{\mathbf{q}})G_0(\vec{\mathbf{k}} - \vec{\mathbf{q}}) \\
+&\frac{q_mP_{mi}(\mathbf{k})k_j\mathsf{Q}_{ij}}{k^2}C_0(\vec{\mathbf{k}} - \vec{\mathbf{q}})G_0(\vec{\mathbf{q}})\Bigg].
\end{aligned} \tag{S17}$$

The internal frequency integrals can be solved exactly

$$\int\frac{d\Omega}{2\pi}G_0(\mathbf{q}, \Omega)G_0(-\mathbf{q}, -\Omega)G_0(\mathbf{k} - \mathbf{q}, \omega - \Omega) = \frac{i}{2\tau_0^{-1}(\mathbf{q})[\omega + i\tau_0^{-1}(\mathbf{q}) + i\tau_0^{-1}(\mathbf{k} - \mathbf{q})]}, \tag{S18}$$

$$\int\frac{d\Omega}{2\pi}G_0(\mathbf{q}, \Omega)G_0(\mathbf{q} - \mathbf{k}, \Omega - \omega)G_0(\mathbf{k} - \mathbf{q}, \omega - \Omega) = \frac{i}{2\tau_0^{-1}(\mathbf{k} - \mathbf{q})[\omega + i\tau_0^{-1}(\mathbf{q}) + i\tau_0^{-1}(\mathbf{k} - \mathbf{q})]}. \tag{S19}$$

Since our interest is in the hydrodynamic regime ($k \to 0$, $\omega \to 0$), we can immediately set $\omega = 0$ in the above

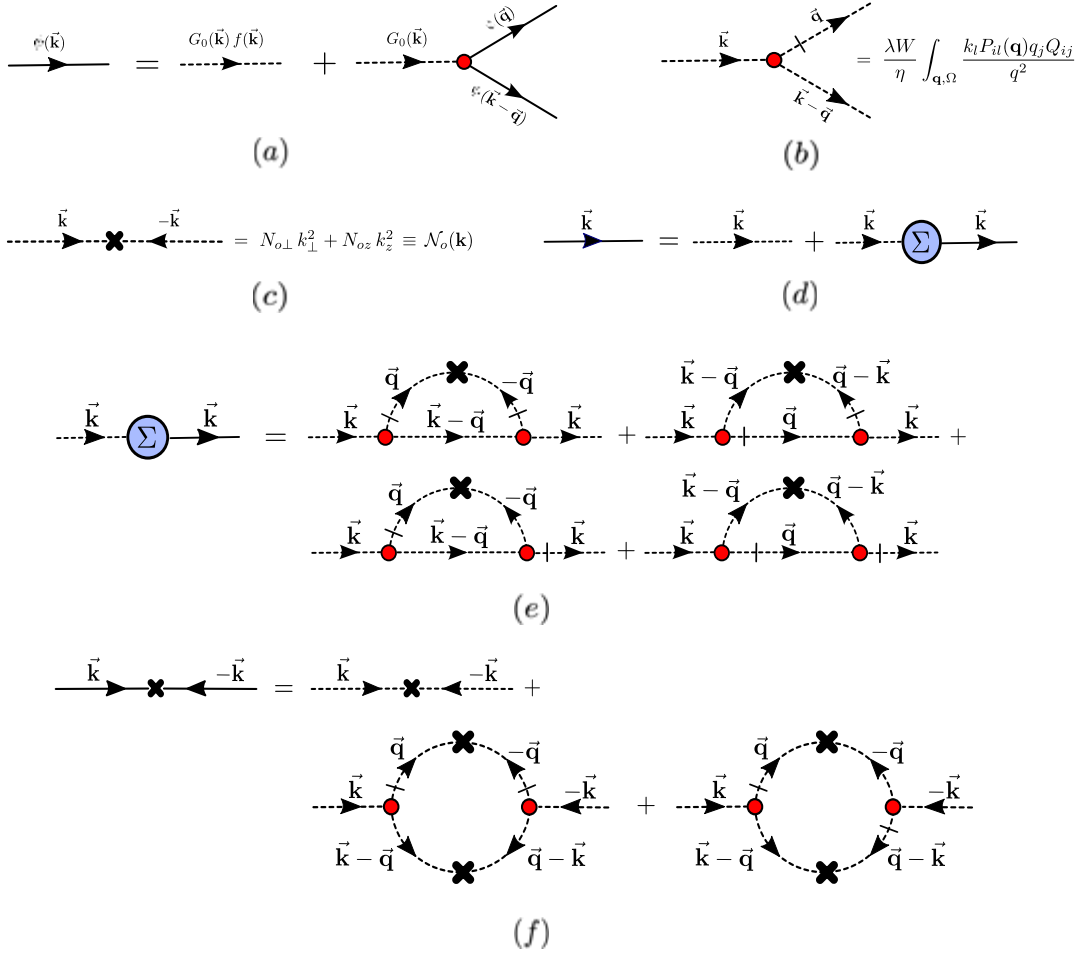

FIG. S1. (a) Diagrammatic representation of Eq. (S11) (b) the three-point vertex of the nonlinearity (c) the two-point noise vertex (d) Diagrammatic representation of the Dyson equation (S16) (e) The self energy at one-loop order (f) One-loop correction to the noise spectrum. Figure adapted with modifications from [1].

equations. Using $P_{il}Q_{ij}q_jk_l = k_zq_z - \frac{q_z^2}{q^2}\mathbf{q}\cdot\mathbf{k}$, the integrals (S17) can be written as

$$\Sigma(\mathbf{k}, 0) = \int \frac{d^d\mathbf{q}}{(2\pi)^d} \frac{k_l}{q^2} \frac{q_z P_{lz}(\mathbf{q})}{\tau_0^{-1}(\mathbf{q}) + \tau_0^{-1}(\mathbf{k}-\mathbf{q})} \left[ \frac{k_m}{(\mathbf{q}-\mathbf{k})^2} \frac{(q_z-k_z)P_{mz}(\mathbf{q}-\mathbf{k})\mathcal{N}_0(\mathbf{k}-\mathbf{q})}{\tau_0^{-1}(\mathbf{k}-\mathbf{q})} - \frac{k_m}{q^2} \frac{q_z P_{mz}(\mathbf{q})\mathcal{N}_0(\mathbf{q})}{\tau_0^{-1}(\mathbf{q})} \right.$$
$$\left. + \frac{q_m k_z P_{mz}(\mathbf{k})}{k^2} \left( \frac{\mathcal{N}_0(\mathbf{k}-\mathbf{q})}{\tau_0^{-1}(\mathbf{k}-\mathbf{q})} - \frac{\mathcal{N}_0(\mathbf{q})}{\tau_0^{-1}(\mathbf{q})} \right) \right]. \quad \text{(S20)}$$

We want $a\partial_\perp^2 \partial_z v_z + b\partial_z^3 v_z$, which with the help of Eq. (S9) can be written as $\frac{aW}{\eta k^4}k_\perp^4 k_z^2 c_k + \frac{bW}{\eta k^4}k_\perp^2 k_z^4 c_k$. Therefore we will focus on the last two terms of the integrand of the $\Sigma(\mathbf{k}, 0)$, which we call $\sigma_2$ that is

$$\sigma_2(\mathbf{k}, \mathbf{q}) \equiv \frac{k_l}{q^2} \frac{q_z P_{lz}(\mathbf{q})}{\tau_0^{-1}(\mathbf{q}) + \tau_0^{-1}(\mathbf{k}-\mathbf{q})} \left[ \frac{q_m k_z P_{mz}(\mathbf{k})}{k^2} \left( \frac{\mathcal{N}_0(\mathbf{k}-\mathbf{q})}{\tau_0^{-1}(\mathbf{k}-\mathbf{q})} - \frac{\mathcal{N}_0(\mathbf{q})}{\tau_0^{-1}(\mathbf{q})} \right) \right], \quad \text{(S21)}$$

because this already has $k^2$ in denominator, first two terms can't produce $k^4$ in the denominator for small $k$. Now defining structure factor

$$S(q) \equiv \frac{\mathcal{N}_0(\mathbf{q})}{\tau_0^{-1}(\mathbf{q})}, \quad \text{(S22)}$$

Eq. ([S21]) can be written as

$$\sigma_2(\mathbf{k}, \mathbf{q}) \equiv \frac{k_l}{q^2} q_z P_{lz}(\mathbf{q}) \frac{q_m k_z P_{mz}(\mathbf{k})}{k^2} \frac{S(\mathbf{k} - \mathbf{q}) - S(\mathbf{q})}{\tau_0^{-1}(\mathbf{q}) + \tau_0^{-1}(\mathbf{k} - \mathbf{q})}. \tag{S23}$$

Symmetrizing Eq. ([S23]) we get

$$\sigma_2(\mathbf{k}, \mathbf{q}) \equiv \frac{k_l}{(\mathbf{q} + \mathbf{k}/2)^2} (q_z + k_z/2) P_{lz}(\mathbf{q} + \mathbf{k}/2) \frac{q_m k_z P_{mz}(\mathbf{k})}{k^2} \frac{S(\mathbf{q} - \mathbf{k}/2) - S(\mathbf{q} + \mathbf{k}/2)}{\tau_0^{-1}(\mathbf{q} + \mathbf{k}/2) + \tau_0^{-1}(\mathbf{q} - \mathbf{k}/2)}. \tag{S24}$$

Defining

$$\mathbb{T}(\mathbf{k}, \mathbf{q}) \equiv \frac{S(\mathbf{q} - \mathbf{k}/2) - S(\mathbf{q} + \mathbf{k}/2)}{\tau_0^{-1}(\mathbf{q} + \mathbf{k}/2) + \tau_0^{-1}(\mathbf{q} - \mathbf{k}/2)}, \tag{S25}$$

Eq. ([S24]) can be written as

$$\sigma_2(\mathbf{k}, \mathbf{q}) \equiv \frac{k_l}{q^2} (q_z + k_z/2) P_{lz}(\mathbf{q} + \mathbf{k}/2) \frac{q_m k_z P_{mz}(\mathbf{k})}{k^2} \left( 1 + \frac{k^2}{4q^2} + \frac{\mathbf{k} \cdot \mathbf{q}}{q^2} \right)^{-1} \mathbb{T}(\mathbf{k}, \mathbf{q}). \tag{S26}$$

Using binomial expansion, we can write

$$\sigma_2(\mathbf{k}, \mathbf{q}) \equiv \frac{q_z + k_z/2}{q^2} \frac{k_z}{k^2} \left( k_z - \frac{(\mathbf{k} \cdot \mathbf{q} + k^2/2)(q_z + k_z/2)}{(\mathbf{q} + \mathbf{k}/2)^2} \right) \left( q_z - \frac{\mathbf{k} \cdot \mathbf{q} k_z}{k^2} \right) \left( 1 - \frac{k^2}{4q^2} - \frac{\mathbf{k} \cdot \mathbf{q}}{q^2} \right) \mathbb{T}(\mathbf{k}, \mathbf{q}) + \dots \tag{S27}$$

Now collecting the terms which give $k^4$ in denominator, we get

$$\sigma_2(\mathbf{k}, \mathbf{q}) \equiv \frac{q_z + k_z/2}{q^2} \frac{k_z}{k^2} \left[ k_z - \frac{(\mathbf{k} \cdot \mathbf{q})(q_z + k_z/2)}{q^2} \left( 1 - \frac{\mathbf{k} \cdot \mathbf{q}}{q^2} \right) \right] \left( -\frac{\mathbf{k} \cdot \mathbf{q} k_z}{k^2} \right) \left( 1 - \frac{\mathbf{k} \cdot \mathbf{q}}{q^2} \right) \mathbb{T}(\mathbf{k}, \mathbf{q}). \tag{S28}$$

Defining $\mu$ as $\mathbf{k}_\perp \cdot \mathbf{q}_\perp = k_\perp q_\perp \mu$, we have

$$\sigma_2(\mathbf{k}, \mathbf{q}) \equiv \frac{q_z + k_z/2}{q^2} \frac{k_z}{k^2} \left[ k_z - \frac{(k_z q_z + k_\perp q_\perp \mu)(q_z + k_z/2)}{q^2} \left( 1 - \frac{\mathbf{k} \cdot \mathbf{q}}{q^2} \right) \right] \left( -\frac{\mathbf{k} . \mathbf{q} k_z}{k^2} \right) \left( 1 - \frac{\mathbf{k} \cdot \mathbf{q}}{q^2} \right) \mathbb{T}(\mathbf{k}, \mathbf{q}). \tag{S29}$$

As $\mathbb{T}(\mathbf{k}, \mathbf{q})$ is odd in $\mathbf{k}$, Taylor expanding it we get

$$\mathbb{T}(\mathbf{k}, \mathbf{q}) = c_1 k_z + c_2 \mathbf{k}_\perp + \dots \text{odd powers} \tag{S30}$$

where

$$\begin{aligned} c_1 &= \left. \frac{\partial \mathbb{T}(\mathbf{k}, \mathbf{q})}{\partial k_z} \right|_{\mathbf{k}=0} = -\frac{4(D_{0\perp} N_{0z} - D_{0z} N_{0\perp}) q_z q_\perp^2}{(D_{0z} q_z^2 + D_{0\perp} q_\perp^2)^3} \\ c_2 &= \left. \frac{\partial \mathbb{T}(\mathbf{k}, \mathbf{q})}{\partial k_\perp} \right|_{\mathbf{k}=0} = \frac{4\mu(D_{0\perp} N_{0z} - D_{0z} N_{0\perp}) q_z^2 q_\perp}{(D_{0z} q_z^2 + D_{0\perp} q_\perp^2)^3}. \end{aligned} \tag{S31}$$

We do not have to expand $\mathbb{T}(\mathbf{k}, \mathbf{q})$ further, as terms with cubic and higher powers will have more power of $k_z$ and/or $\mathbf{k}_\perp$ in the numerator than required, which we have shown will be irrelevant in RG sense. Now we can pull out the terms of type $\frac{k_z^2}{k^4} \frac{k_\perp^3}{k^4}, \frac{k_z^3}{k^4} \frac{k_z^2}{k^4}, \frac{k_\perp}{k^4} \frac{k_z^4}{k^4}$, and multiply by obtained expansion of $\mathbb{T}(\mathbf{k}, \mathbf{q})$ to get required FIM terms. Using Mathematica we find,

$$a = \frac{\lambda^2 W}{\eta} \int \frac{d^d \mathbf{q}}{(2\pi)^d} \left( \frac{-2q_\perp^4 q_z^4}{q^6} \right) \mu^4 r(\mathbf{q}), \tag{S32}$$

and

$$b = \frac{\lambda^2 W}{\eta} \int \frac{d^d \mathbf{q}}{(2\pi)^d} \left( \frac{6(q_\perp^4 q_z^4 - q_\perp^2 q_z^6)}{q^6} + \frac{2(2q_\perp^2 q_z^4 - q_\perp^4 q_z^2)}{q^4} - \frac{q_\perp^2 q_z^2}{2q^2} \right) \mu^2 r(\mathbf{q}), \tag{S33}$$

where,

$$r(\mathbf{q}) = \frac{4(D_{0\perp}N_{0z} - D_{0z}N_{0\perp})}{(D_{0z}q_z^2 + D_{0\perp}q_\perp^2)^3}. \tag{S34}$$

Note that the factor $r(\mathbf{q})$ in the $a$ and $b$ integrals above is zero if the mobility and noise are constrained by a Fluctuation Dissipation Relation, i.e., when $N_{0z}/N_{0\perp} = D_{0z}/D_{0\perp}$. Hence, the FIM terms are only possible in a nonequilibrium system.

## III. SYSTEM WITH PLANAR CONFINEMENT

In this section, we show that even when the system is confined to a quasi-2D geometry between planar walls parallel to the $xz$ plane, with a small $y$ separation, the active contribution to the particle current still leads to condensation through a mechanism similar to that in the unbounded 3D bulk case. We will follow an approach similar to that used for active nematic suspensions on a substrate (see e.g. [2] and the references therein).

We assume here that the two parallel confining plates impose no-slip, no-penetration boundary conditions on the velocity field, no-flux boundary condition on the concentration field, and we have planar anchoring for swimmers inherent throughout the system. Considering the system is confined along the $y$-direction by walls separated by a distance $h$, the no-slip boundary conditions at the walls constrain the longest permissible wavelength in the $y$-direction to $\sim h$, so we will use a lubrication approximation in which terms in the equations of motion at higher orders in $\epsilon = h/L$ are discarded for in-plane ($xz$-plane) length scale $L \gg h$. The lubrication approximation implies $\partial_y \gg \partial_x, \partial_z$ and hence, $\partial_y^2 \gg \partial_x^2, \partial_z^2$ for $L \gg h$. In the zeroth order of $\epsilon$, for 3D fluid velocity filed $\bar{\mathbf{u}}$, the incompressibility of fluid field reads $\partial_y \bar{u}_y = 0$ implying $\bar{u}_y$ is independent of $y$ and due to no-penetration boundary condition in the $y$ direction at the walls, we have $\bar{u}_y = 0$. So for simplicity, let's assume a Poiseuille profile for the flow field satisfying no-slip boundary condition at the walls ($y = 0$ and $y = h$)

$$\bar{\mathbf{u}}_{xz}(y) = \frac{4}{h^2}\bar{\mathbf{u}}_{0xz}(hy - y^2), \tag{S35}$$

where $\bar{\mathbf{u}}_{0xz}$ is the mid-plane velocity field parallel to the walls. So the $y$-averaged velocity field in the $xz$ plane, $\mathbf{u} = \frac{2}{3}\bar{\mathbf{u}}_{0xz}$ and the $y$-averaged viscous force density in the Stokes equation Eq. (S5) becomes

$$\frac{1}{h}\int_0^h \eta\nabla^2\bar{\mathbf{u}}_{xz}(y)dy = -\Gamma\mathbf{u} + \mathcal{O}\left(\epsilon^2\right), \tag{S36}$$

where $\Gamma = 12\eta/h^2$ is the effective damping coefficient. Note that any other flow profile satisfying the imposed boundary conditions above, even if it is not of the Poiseuille type, can only affect the numerical value of the coefficient $\Gamma$.

So, the effective ($y$-averaged) Stokes equation for the two-dimensional velocity field is

$$\Gamma\mathbf{u} = -\nabla\mathcal{P} - W\hat{\mathbf{z}}\partial_z c, \quad \nabla\cdot\mathbf{u} = \mathbf{0}. \tag{S37}$$

Similarly, after averaging over the thin direction, the FIM current becomes of the form

$$\mathbf{J^u} \sim \mathbf{A}\nabla\nabla\bar{\mathbf{u}} \sim \mathbf{Bu} = \frac{c_1}{h^2}u_x\hat{\mathbf{x}} + \frac{c_2}{h^2}u_z\hat{\mathbf{z}}, \tag{S38}$$

where $\mathbf{A}$, $\mathbf{B}$ are fourth-rank and second-rank tensor, respectively. Note that the system considered here is liberated from Galilean invariance, hence currents linear in components $\mathbf{u}$ and independent of $c$ are possible satisfying $\hat{\mathbf{z}} \leftrightarrow -\hat{\mathbf{z}}$ symmetry of the system.

Applying a similar analysis with the lubrication approximation and the no-flux boundary condition at the walls leads to the term of form $\mathbf{Q}\cdot\nabla(\nabla^2 c)$, which was disregarded in a gradient-expansion treatment for the unconfined 3D case, contributing in the same order as the active stress in Eq. (S37) when averaged over the $y$-direction. Since this term do not introduce any qualitatively new features, we can redefine $W$ differing from the force dipole strength defined for the bulk system.

Solving the coupled equations of Eq. (S37) and the continuity equation for $c$ with active particle currents from Eq. (S38) and neglecting advection, as done for the 3D case, we obtain in Fourier space: $\partial_t c_k = -D(\theta)k^2 c_k$ where $\theta$ is the angle between the wavevector $\mathbf{k}$ and the $\hat{\mathbf{z}}$ axis, with

$$D(\theta) = \left(D_z - \frac{(c_1 - c_2)W}{2\eta}\sin^2\theta\right)\cos^2\theta + \left(D_x - \frac{(c_1 - c_2)W}{2\eta}\cos^2\theta\right)\sin^2\theta. \tag{S39}$$

Note that the redefinition of $W$ in Eq. (S37), due to broken Galilean invariance and possible contribution from higher order active stress defined above, makes $W$ another phenomenological parameter in the problem, hence, unlike for the bulk case, the obtained diffusivity shift $\sim \frac{(c_1-c_2)W}{\eta}$ may not always be positive for the Stokesian swimmers. Although the migration velocity estimate in [3] still implies $c_1 > 0$ and $c_2 < 0$ for this quasi-2D case, the redefinition of $W$ does not suggest that their respective contributions lead to instability in extensile and contractile systems, as observed in the bulk system. So condensation in the quasi-2D system can be achieved by varying values of $c_1$, $c_2$ and $W$ such that $\frac{(c_1-c_2)W}{2\eta} \geq D_{z/x} > 0$.

## IV. LINEAR STABILITY ANALYSIS WITH UNSTEADY STOKES EQUATION

In this section, we study linear stability of state with homogeneous concentration $c = c_0$, $\mathbf{u} = 0$ for small perturbation $\delta c$ and $\delta \mathbf{u} = (\mathbf{u}_\perp, u_z)$. Considering monochromatic perturbation $(\delta c, \delta \mathbf{u}) = (\hat{\delta c}, \hat{\delta \mathbf{u}}) e^{i(\mathbf{k}\cdot\mathbf{r} - \omega t)}$, where $\mathbf{k} = (\mathbf{k}_\perp \cdot \perp + k_z \hat{\mathbf{z}}) = k(\sin\theta \perp + \cos\theta \hat{\mathbf{z}})$, up to linear order, the continuity equation for $c$ and the unsteady Stokes equation corresponding to Eq. (S5) becomes

$$(-i\omega + D_\perp k_\perp^2 + D_z k_z^2)\hat{\delta c} = i(ak_\perp^2 + bk_z^2)k_z \hat{\delta u}_z, \tag{S40}$$

$$(-i\rho\omega + \eta k^2)\hat{\delta u}_z = -i\frac{Wk_\perp^2}{k^2}k_z \hat{\delta c}. \tag{S41}$$

From Eqs. (S40) and (S41), we have the dispersion relation

$$\frac{2\omega_\pm}{ik^2} = -\left(D_\perp \sin^2\theta + D_z \cos^2\theta + \frac{\eta}{\rho}\right) \pm \left[\left(D_\perp \sin^2\theta + D_z \cos^2\theta - \frac{\eta}{\rho}\right)^2 + \left(\frac{aW}{\rho}\sin^2\theta + \frac{bW}{\rho}\cos^2\theta\right)\sin^2 2\theta\right]^{1/2} \tag{S42}$$

For Stokesian swimmers, $aW$ and $bW$ are always positive, as discussed in the main text. This implies $\text{Im}(RHS) = 0$ in Eq. (S42) for all possible parameter values, indicating that no oscillatory modes can exist in the system.

For $\text{Re}(RHS) > 0$ in Eq. (S42) we have instability/condensation, which is possible *iff*

$$\left(D_\perp \sin^2\theta + D_z \cos^2\theta\right)\frac{4\eta}{\rho} < \left(\frac{aW}{\rho}\sin^2\theta + \frac{bW}{\rho}\cos^2\theta\right)\sin^2 2\theta$$
$$\Rightarrow \left(D_\perp - \frac{aW}{4\eta}\sin^2 2\theta\right)\sin^2\theta + \left(D_z - \frac{bW}{4\eta}\sin^2 2\theta\right)\cos^2\theta < 0 \tag{S43}$$

The $D(\theta)$ expression obtained above is exactly same as the steady Stokes case. Hence, although the inertia changes diffusive relaxation rate, as we can see from Eq. (S42), it does not affect the condensation.

## V. NUMERICAL SIMULATION DETAILS

In this section we briefly discuss the numerical simulation techniques used in our study. The over-damped particle evolution equation in our simulation includes, in addition to the long-range hydrodynamic interaction, a background of Gaussian white noise

$$\frac{d\mathbf{r}_\alpha}{dt} = \mathbf{u}(\mathbf{r}_\alpha(t), t) + \sqrt{2D}\boldsymbol{\zeta}_\alpha(t), \tag{S44}$$

where,

$$\langle \boldsymbol{\zeta}_\alpha(t) \otimes \boldsymbol{\zeta}_\beta(t') \rangle = \delta(t - t')\,\delta_{\alpha\beta}\,\mathbf{I}, \tag{S45}$$

where $\mathbf{u}(\mathbf{r}_\alpha(t), t)$ is the fluid velocity field at the particle position $\mathbf{r}_\alpha$ at time $t$. For a collection of force dipoles in bulk permanently aligned along the $\hat{\mathbf{z}}$-direction in an incompressible fluid medium with dipole strength $W$, with assigned nematic order parameter $\mathbf{Q} = \hat{\mathbf{z}}\hat{\mathbf{z}} - \frac{\mathbf{I}}{3}$, the corresponding Stokes equation is given by

$$\eta\nabla^2\mathbf{u} = \nabla\mathcal{P} + W\mathbf{Q}\cdot\nabla\sum_\alpha \delta(\mathbf{r} - \mathbf{r}_\alpha), \quad \nabla\cdot\mathbf{u} = 0. \tag{S46}$$

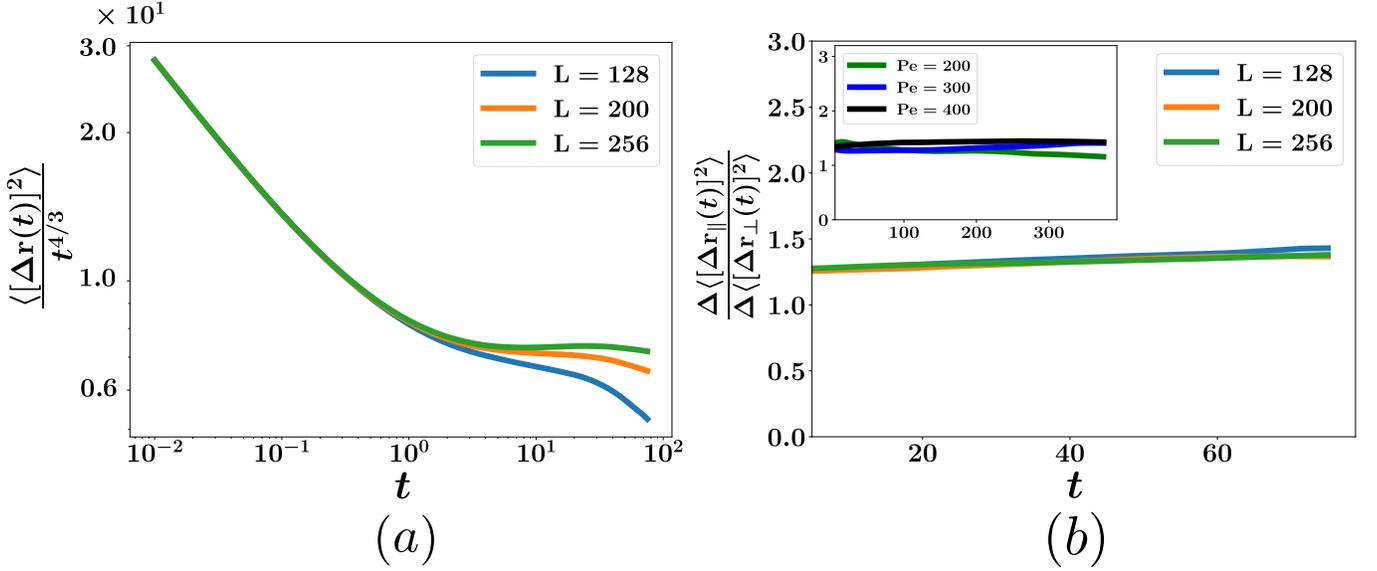

FIG. S2. (a) Mean-Square Displacement (MSD) divided by proposed asymptotic superdiffusive contributions to MSD (i.e., $L^2 \sim \tau^{\frac{4}{3}}$ for 3D) for different box sizes at Pe = 1000 and a concentration of 0.0005. The plot shows a growing plateau region with increasing box size, demonstrating the finite-size effect in the simulation. (b) Ratio of filtered MSD components for different system sizes, representing anisotropic diffusion in the system. The results show enhanced diffusion in the parallel ($\parallel / \hat{z}$) direction compared to the perpendicular direction. (*inset*) Shows the ratio of filtered MSD components for different Péclet numbers for $L = 200\ell$. See the movie mov1-convex-hull.mov in the Supplemental Material, which animates the convex hull of the particle cloud initially starting from the center of the box, highlighting the anisotropic diffusion in the system.

Defining the Fourier transform as

$$\mathbf{u_k} = \mathcal{F}[\mathbf{u(r)}] = \int d^3\mathbf{r} \ e^{-i\mathbf{k}\cdot\mathbf{r}} \ \mathbf{u(r)}, \tag{S47}$$

in Fourier space, the solution to the Stokes Eq. (S46) becomes

$$\mathbf{u_k} = \frac{-iW}{\eta k^2} \ \mathbf{\Pi_k} \cdot \mathbf{Q} \cdot \mathbf{k} \sum_\alpha e^{-i\mathbf{k}\cdot\mathbf{r_\alpha}}, \quad \mathbf{\Pi_k} = \mathbf{I} - \hat{\mathbf{k}}\hat{\mathbf{k}}. \tag{S48}$$

Choosing unit cell size of the system $\ell$ as the length scale and $\ell^2/D$ as the timescale, in non-dimensional units, the particle equation of motion becomes

$$\frac{d\mathbf{r_\alpha}}{dt} = \mathbf{u(r_\alpha)} + \sqrt{2}\boldsymbol{\zeta}_\alpha(t), \tag{S49}$$

and the flow field in real space is given by

$$\mathbf{u(r)} = \text{Pe} \sum_{\mathbf{k}} \sum_\alpha \frac{ie^{i\mathbf{k}\cdot(\mathbf{r}-\mathbf{r_\alpha})}}{k^4} \left[ k_z^2(k_x\hat{\mathbf{x}} + k_y\hat{\mathbf{y}}) - (k_x^2 + k_y^2) k_z\hat{\mathbf{z}} \right], \tag{S50}$$

where we define the Péclet number, Pe $\equiv \frac{W}{\eta D \ell}$.

The $\mathbf{u(r_\alpha)}$ in Eq. (S49) is the interpolated $\mathbf{u(r)}$ at the particle positions $\mathbf{r_\alpha}$. We solve the Stokes equation using Fourier spectral method and employ a 4-point interpolation algorithm [4] to interpolate velocity field at the off-grid Lagrangian particle position from Eulerian fluid velocity field. Note that the interpolation algorithm along with spatial discretization used here defines a particle size $\ell$ in the problem[1, 5].

Particles are initially randomly dispersed throughout the periodic domain and follow Eqs. (S49) and (S50). We use Euler-Maruyama integration scheme for the time matching with a time-step $\delta t = 0.0005\ell^2/D$. The particle position data is stored ignoring initial $2 \times 10^4$ iterations and after every 20 iterations for a well-converged statistics.

The Mean Square Displacement (MSD) is calculated using a moving time average of particle position data obtained

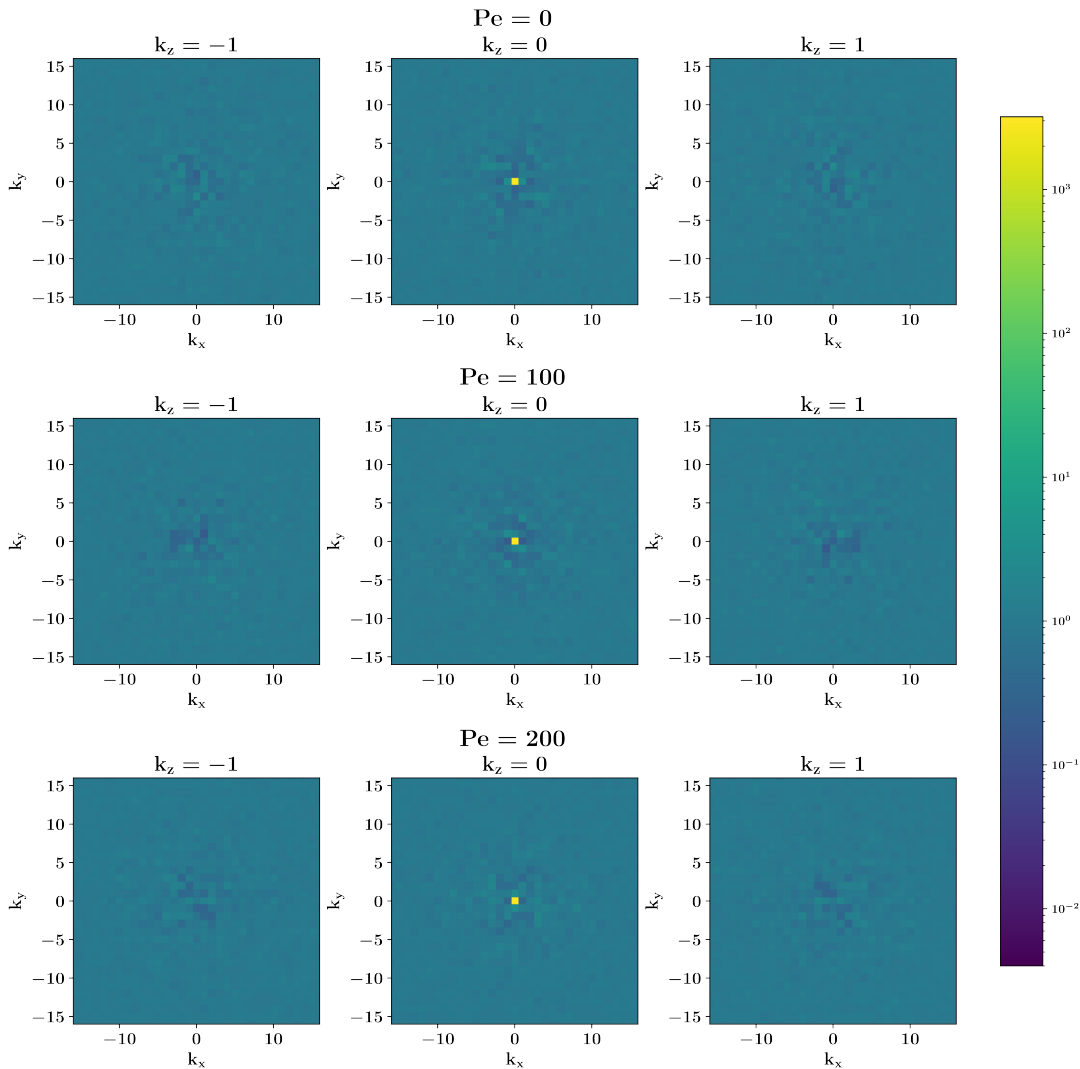

FIG. S3. Heat map of static structure factor (in log scale) for different Péclet numbers demonstrates that the static structure factor is independent of wavevector, similar to its passive counterpart (Pe = 0 in the Figure) thereby verifying the roughness exponent $\chi = -\frac{3}{2}$ in 3D. The integer k values in the plot correspond to Fourier mode indices. See the movie mov2-particle-dyn.mov in the Supplemental Material, animating particle dynamics for different Péclet numbers.

from a single simulation with given parameter values, run for a long duration of $10^6$ iterations. Fig. S2 (a) illustrates the finite-size scaling of the simulation, showing a plateau region that grows with increasing box size. Although we used a relatively high Péclet number in this plot to highlight the superdiffusive region of the MSD, the very small time-step ensures that the Euler scheme remains valid. We considered particle positions up to time steps where particles, on average, do not cross half of the box size to maintain good statistical accuracy with periodic boundary condition. To better extract the asymptotic superdiffusive behavior, without going to high Péclet number, we substract out the exactly known diffusive components from the calculated MSD. So, For a spatial dimension $d$, the filtered MSD is $\Delta\langle\Delta\mathbf{r}^2(t)\rangle = \langle\Delta\mathbf{r}^2(t)\rangle - 2dt$. Fig. S2 (b) displays the anisotropic diffusion for different box sizes, with the (*inset*) showing the results for different Péclet numbers.

Fig. S3 show the heatmap of static structure factor. The static structure factor plotted here is obtained by averaging over last 10000 iterations, ensuring that no transient behavior is present and that the data is from a steady state. The heatmap indicates that the static structure factor is independent of the wavevector, similar to its passive counterpart, thereby confirming the predicted roughness exponent $\chi = -\frac{3}{2}$ in 3D.

## VI. EARLY TIME BALLISTIC BEHAVIOR

In this section, we give an explanation for the observed early-time ballistic behavior in the system. Integrating the particle position evolution Eq. (S44).

$$\Delta \mathbf{r}_\alpha(t) \equiv \mathbf{r}_\alpha(t) - \mathbf{r}_\alpha(0) = \int_0^t ds\, \mathbf{u}(\mathbf{r}_\alpha(s), s) + \sqrt{2D} \int_0^t ds\, \boldsymbol{\zeta}_\alpha(s). \tag{S51}$$

Ignoring the Brownian noise, the MSD of the particles is given by

$$\Delta \langle [\Delta \mathbf{r}(t)]^2 \rangle = \int_{s=0}^t ds \int_{s'=0}^t ds' \, \langle \mathbf{u}(\mathbf{r}_\alpha(s), s) \cdot \mathbf{u}(\mathbf{r}_\alpha(s'), s') \rangle. \tag{S52}$$

We have omitted the particle index $\alpha$ on the left-hand side of Eq. (S52), as the notation $\langle \rangle$ represents average over all the particles in addition to the moving-time average. Using Eq. (S48), Particle Velocity Auto Correlation Function (PVACF) becomes

$$\langle \mathbf{u}(\mathbf{r}_\alpha(s), s) \cdot \mathbf{u}(\mathbf{r}_\alpha(s'), s') \rangle = \left( \frac{W}{\eta} \right)^2 \sum_{\mathbf{k}} \sum_{\mathbf{q}} \sum_\beta \sum_\gamma \left\langle e^{i \left[ \mathbf{k} \cdot (\mathbf{r}_\alpha(s) - \mathbf{r}_\beta(s)) + \mathbf{q} \cdot (\mathbf{r}_\alpha(s') - \mathbf{r}_\gamma(s')) \right]} \right\rangle$$
$$\frac{\left[ k_z^2 (k_x \hat{\mathbf{x}} + k_y \hat{\mathbf{y}}) - (k_x^2 + k_y^2) k_z \hat{\mathbf{z}} \right] \cdot \left[ q_z^2 (q_x \hat{\mathbf{x}} + q_y \hat{\mathbf{y}}) - (q_x^2 + q_y^2) q_z \hat{\mathbf{z}} \right]}{k^4 \, q^4}, \tag{S53}$$

where for simplicity, we have set $\Delta k, \Delta q = 1$.

In the steady state, the system's static structure factor becomes featureless, similar to its passive counterpart, as observed in the previous section (Sec. V, Fig. S3). This leads to two consequences for Eq. (S53), (1) the contribution to the PVACF from its imaginary part vanishes at all times in the thermodynamic limit [6, 7] and (2) at very early times, the dynamics is primarily governed by its equal-time 'self' (i.e. single-particle) component, as no correlation exist among $\mathbf{r}_\alpha$, $\mathbf{r}_\beta$ and $\mathbf{r}_\gamma$ [8]. So at the early time limit, we can write Eq. (S52) as

$$\Delta \langle [\Delta \mathbf{r}(t)]^2 \rangle \overset{t \to 0}{\sim} \left( \frac{W}{\eta} \right)^2 t^2 \sum_{\mathbf{k}} \sum_{\mathbf{q}} \sum_{\beta = \alpha} \langle \cos \left[ (\mathbf{k} + \mathbf{q}) \cdot (\mathbf{r}_\alpha(0) - \mathbf{r}_\beta(0)) \right] \rangle$$
$$\frac{\left[ k_z^2 (k_x \hat{\mathbf{x}} + k_y \hat{\mathbf{y}}) - (k_x^2 + k_y^2) k_z \hat{\mathbf{z}} \right] \cdot \left[ q_z^2 (q_x \hat{\mathbf{x}} + q_y \hat{\mathbf{y}}) - (q_x^2 + q_y^2) q_z \hat{\mathbf{z}} \right]}{k^4 \, q^4}. \tag{S54}$$

In the continuum limit, replacing particle summation by integration over all space we get,

$$\Delta \langle [\Delta \mathbf{r}(t)]^2 \rangle \overset{t \to 0}{\sim} \left( \frac{W}{\eta} \right)^2 t^2 c \sum_{\mathbf{k}} \sum_{\mathbf{q}}$$
$$\frac{\left[ k_z^2 (k_x \hat{\mathbf{x}} + k_y \hat{\mathbf{y}}) - (k_x^2 + k_y^2) k_z \hat{\mathbf{z}} \right] \cdot \left[ q_z^2 (q_x \hat{\mathbf{x}} + q_y \hat{\mathbf{y}}) - (q_x^2 + q_y^2) q_z \hat{\mathbf{z}} \right]}{k^4 \, q^4}, \tag{S55}$$

where $c$ is the concentration. As there is a small-scale cutoff $\ell$ in our simulation, the sum over wave vectors in Eq. (S55) will yield a finite value for the coefficient of $t^2$.

---



\end{document}